\documentclass[proceedings]{JHEP3}
\usepackage{amsfonts}
\usepackage{amsmath}
\usepackage{epsfig,multicol}

\newbox\mybox

\newcommand\fverb{\setbox\mybox=\hbox\bgroup\verb}
\newcommand\fverbdo{\egroup\medskip\noindent\fbox{\unhbox\mybox}\ }
\newcommand\fverbit{\egroup\item[\fbox{\unhbox\mybox}]}
\conference{The Ising quantum spin chain in an imaginary field}
\abstract{We investigate a lattice version of the
Yang-Lee model which is characterized by a non-Hermitian quantum
spin chain Hamiltonian. We propose a new way to implement
$\mathcal{PT}$-symmetry on the lattice, which serves to guarantee the reality of the
spectrum in certain regions of values of the coupling constants.
In that region of unbroken $\mathcal{PT}$-symmetry we construct
a Dyson map, a metric operator and find the Hermitian counterpart
of the Hamiltonian for small values of the number of sites, both
 exactly and perturbatively. Besides the standard perturbation
 theory about the Hermitian part of the Hamiltonian, we also carry
 out an expansion in the second coupling constant of the model.
Our constructions turns out to be unique with the sole assumption
 that the Dyson map is Hermitian.
Finally we compute the magnetization of the chain in the $z$ and
$x$ direction.}

\title{A spin chain model with non-Hermitian interaction: The Ising quantum
spin chain in an imaginary field}
\author{Olalla A.~Castro-Alvaredo and Andreas Fring \\
Centre for Mathematical Science, City University London, \\
Northampton Square, London EC1V 0HB, UK\\
E-mail: O.Castro-alvaredo@city.ac.uk, A.Fring@city.ac.uk}

\input{tcilatex}

\begin{document}

\section{Introduction}

It is known for about thirty years that ordinary second order phase
transitions can be described by the Yang-Lee model \cite{YL,LY,Fisher}. This
model admits a quantum field theoretical description in form of a
Landau-Ginzburg Hamiltonian for a scalar field $\phi $ with an additional $%
\phi ^{3}$-interaction and a term linear in the scalar field with an
imaginary coupling constant. The model has been identified \cite%
{Cardy:1985yy} as a perturbation of the $\mathcal{M}_{5,2}$-model in the $%
\mathcal{M}_{p,q}$-series of minimal conformal field theories \cite{BPZ}. It
is the simplest non-unitary model in this infinite class of models, which
are all characterized by the condition $p-q>1$ and whose corresponding
Hamiltonians are all expected to be non-Hermitian.

Here we shall investigate a discretised lattice version of the Yang-Lee
model considered by von Gehlen \cite{gehlen1,gehlen2}, which is an Ising
quantum spin chain in the presence of a magnetic field in the $z$-direction
as well as a longitudinal imaginary field in the $x$-direction. The
corresponding Hamiltonian for a chain of length $N$ is given by 
\begin{equation}
H(\lambda ,\kappa )=-\frac{1}{2}\sum_{j=1}^{N}(\sigma _{j}^{z}+\lambda
\sigma _{j}^{x}\sigma _{j+1}^{x}+i\kappa \sigma _{j}^{x}),\qquad \lambda
,\kappa \in \mathbb{R}.  \label{H}
\end{equation}%
It acts on a Hilbert space of the form $(\mathbb{C}^{2})^{\otimes N}$ where
we employed the standard notation for the $2^{N}\times 2^{N}$-matrices $%
\sigma _{i}^{x,y,z}=\mathbb{I\otimes I\otimes \ldots \otimes }\sigma
^{x,y,z}\otimes \ldots \otimes \mathbb{I\otimes I}$ with Pauli matrices
describing spin 1/2 particles 
\begin{equation}
\sigma ^{x}=\left( 
\begin{array}{cc}
0 & 1 \\ 
1 & 0%
\end{array}%
\right) ,\qquad \sigma ^{y}=\left( 
\begin{array}{cc}
0 & -i \\ 
i & 0%
\end{array}%
\right) ,\qquad \sigma ^{z}=\left( 
\begin{array}{cc}
1 & 0 \\ 
0 & -1%
\end{array}%
\right) ,  \label{pauli}
\end{equation}%
as $i$th factor acting on the site $i$ of the chain. Their commutation
relations are direct sums of su(2) algebras 
\begin{equation}
\lbrack \sigma _{j}^{x},\sigma _{k}^{y}]=2i\sigma _{j}^{z}\delta _{jk},\quad
\lbrack \sigma _{j}^{z},\sigma _{k}^{x}]=2i\sigma _{j}^{y}\delta _{jk},\quad
\lbrack \sigma _{j}^{y},\sigma _{k}^{z}]=2i\sigma _{j}^{x}\delta _{jk},\quad 
\text{with }j,k=1,\ldots ,N  \label{com}
\end{equation}%
A further real parameter $\beta $ may be introduced into the model by
allowing different types of boundary conditions $\sigma _{N+1}^{x,y,z}=\beta
\sigma _{1}^{x,y,z}$, albeit here we will only consider the case of periodic
boundary conditions and take $\beta =1$.

Since all Pauli matrices are Hermitian it is obvious that $H(\lambda ,\kappa
)$ is non-Hermitian 
\begin{equation}
H^{\dagger }(\lambda ,\kappa )=H(\lambda ,-\kappa )\neq H(\lambda ,\kappa ).
\end{equation}
This poses immediately two questions: First of all, is the spectrum still
real, despite the fact that the vital property of Hermiticity which
guarantees this is given up and second is it still possible to formulate a
meaningful quantum mechanical description associated to this type of
Hamiltonians? These issues have attracted a considerable amount of attention
in the last ten years, since the seminal paper by Bender and Boettcher \cite%
{BB} and meanwhile many satisfying answers have been found to most of them;
for recent reviews see \cite{Rev1,Rev2,Rev3}.

Our manuscript is organised as follows: In section 2 we present various
alternatives about how $\mathcal{PT}$-symmetry can be implemented for
quantum spin chains. In section 3 we establish our notation and recall some
of the well known facts concerning a consistent quantum mechanical framework
for $\mathcal{PT}$-symmetric systems. We analyze the model (\ref{H}) in
section 4 and section 5, where the former is devoted to non-perturbative and
the latter to perturbative results. In section 6 we compute the
magnetization for the model (\ref{H}) and we state our conclusions in
section 7.

\section{$\mathcal{PT}$-symmetry for spin chains}

Preceding the above mentioned recent activities von Gehlen found numerically 
\cite{gehlen1,gehlen2} that for certain values of the dimensional parameters 
$\lambda $ and $\kappa $ the eigenvalues for $H(\lambda ,\kappa )$ are all
real, whereas for the remaining values they occur in complex conjugate
pairs. He provided an easy explanation for this feature: Acting adjointly on
the Hamiltonian with a spin rotation operator 
\begin{equation}
\mathcal{R}=e^{\frac{i\pi }{4}S_{z}^{N}}=\prod\limits_{i=1}^{N}\frac{1}{%
\sqrt{2}}(\mathbb{I}+i \sigma ^{z})_{i},\quad \text{with }\quad
S_{z}^{N}=\sum\limits_{i=1}^{N}\sigma _{i}^{z},\quad \mathbb{I}=\left( 
\begin{array}{cc}
1 & 0 \\ 
0 & 1%
\end{array}%
\right) ,
\end{equation}%
has the effect of rotating the spins at each site clockwise by $\pi /2$ in
the $xy$-plane, such that the corresponding map acts as $\mathcal{R}:(\sigma
_{i}^{x},\sigma _{i}^{y},\sigma _{i}^{z})\rightarrow (-\sigma
_{i}^{y},\sigma _{i}^{x},\sigma _{i}^{z})$. The resulting Hamiltonian is a $%
2^{N}\times 2^{N}$ non-symmetric matrix with real entries given by 
\begin{equation}
\hat{H}(\lambda ,\kappa )=\mathcal{R}H(\lambda ,\kappa )\mathcal{R}^{-1}=-%
\frac{1}{2}\sum_{i=1}^{N}(\sigma _{i}^{z}+\lambda \sigma _{i}^{y}\sigma
_{i+1}^{y}-i\kappa \sigma _{i}^{y}).  \label{u}
\end{equation}%
Its eigenvalues and those of $H(\lambda ,\kappa )$ are therefore either all
real or occur in complex conjugate pairs. This is precisely the well known
behaviour one finds when $H(\lambda ,\kappa )$ is symmetric with respect to
an anti-linear operator \cite%
{EW,Bender:1998ke,Bender:2002vv,SW,Bender:2004sa}, which as mentioned above
has recently attracted a lot of attention. In quantum mechanical or field
theoretical models the anti-linear operator is commonly taken to be the $%
\mathcal{PT}$-operator, which carries out a simultaneous parity
transformation $\mathcal{P}:x\rightarrow -x$ and time reversal $\mathcal{T}%
:t\rightarrow -t$. When acting on complex valued functions the anti-linear
operator $\mathcal{T}$ \ is understood to act as complex conjugation. Real
eigenvalues are then found for unbroken $\mathcal{PT}$-symmetry, meaning
that both the Hamiltonian \textit{and} the eigenfunctions remain invariant
under $\mathcal{PT}$-symmetry, whereas broken $\mathcal{PT}$-symmetry leads
to complex conjugate pairs of eigenvalues.

We will now argue that $\mathcal{PT}$-symmetry on the lattice can be
interpreted in various ways. One may for instance reflect the chain across
its midpoint via the map $\mathcal{P}^{\prime }:\sigma
_{i}^{x,y,z}\rightarrow \sigma _{N+1-i}^{x,y,z}$ as suggested by Korff and
Weston \cite{Korff:2007qg} and used thereafter in \cite{CKPT}. It is obvious
that the Hamiltonian (\ref{H}) is invariant with regard to this symmetry.
However, when keeping the interpretation of $\mathcal{T}$ as a complex
conjugation, and thus ensuring that the $\mathcal{P}^{\prime }\mathcal{T}$%
-operator is anti-linear, one easily observes that this type of
transformation does not leave the Hamiltonian (\ref{H}) invariant, i.e. we
have $\left[ \mathcal{P}^{\prime }\mathcal{T},H\right] \neq 0$.

Therefore we need to implement $\mathcal{PT}$-symmetry in a different way
for $H(\lambda ,\kappa )$ to be able to analyze its properties along the
lines proposed in \cite{EW,Bender:1998ke,Bender:2002vv,SW,Bender:2004sa}. We
propose here that one carries out a parity transformation at each individual
site and reflect every spin for instance in the $xy$-plane on $y=-x$. This
is obviously achieved by $\mathcal{R}^{2}$. As $\mathcal{R}%
^{4}=\prod_{i=1}^{N}(-\mathbb{I})_{i}=(-1)^{N}\mathbb{I}^{\otimes N}$ and
not the desired identity operator, we take here 
\begin{equation}
\mathcal{P}=-i\mathcal{R}^{2}=e^{\frac{i\pi }{2}(S^{z}-\mathbb{I}%
)}=\prod_{i=1}^{N}\sigma _{i}^{z},\quad \text{with }\quad \mathcal{P}^{2}=%
\mathbb{I}^{\otimes N},  \label{PPP}
\end{equation}
as our parity operator. Consequently this transformation acts as 
\begin{equation}
\mathcal{P}:(\sigma _{i}^{x},\sigma _{i}^{y},\sigma _{i}^{z})\rightarrow
(-\sigma _{i}^{x},-\sigma _{i}^{y},\sigma _{i}^{z}).  \label{eff}
\end{equation}
Thus with $\mathcal{T}$ being the usual complex conjugation, which acts on
the Pauli matrices as 
\begin{equation}
\mathcal{T}:(\sigma _{i}^{x},\sigma _{i}^{y},\sigma _{i}^{z})\rightarrow
(\sigma _{i}^{x},-\sigma _{i}^{y},\sigma _{i}^{z}),  \label{teff}
\end{equation}
we have identified an anti-linear operator constituting a symmetry of the
Hamiltonian (\ref{H}) 
\begin{equation}
\left[ \mathcal{PT},H\right] =0.  \label{PTH}
\end{equation}
This operator provides more information than the transformation (\ref{u}),
because we have now in addition a concrete criterium, which distinguishes
the regimes of real and complex eigenvalues. We can precisely separate the
two domains $U_{\mathcal{PT}}$ and $U_{b\mathcal{PT}}$ in the parameter
space of $\lambda $ and $\kappa $ defined by the action on the eigenstates $%
\Phi (\lambda ,\kappa )$ of $H(\lambda ,\kappa )$%
\begin{equation}
\mathcal{PT}\Phi (\lambda ,\kappa )\left\{ 
\begin{array}{l}
=\Phi (\lambda ,\kappa )\text{ \quad for }(\lambda ,\kappa )\in U_{\mathcal{%
PT}} \\ 
\neq \Phi (\lambda ,\kappa )\text{ \quad for }(\lambda ,\kappa )\in U_{b%
\mathcal{PT}.}%
\end{array}
\right.  \label{U}
\end{equation}
According to the general reasoning provided in \cite%
{EW,Bender:1998ke,Bender:2002vv,SW,Bender:2004sa}, simultaneous
eigenfunctions of $\mathcal{PT}$ and $H(\lambda ,\kappa )$, that is for $%
(\lambda ,\kappa )\in U_{\mathcal{PT}}$, are then associated with real
eigenvalues whereas in the regime of broken $\mathcal{PT}$-symmetry, that is 
$(\lambda ,\kappa )\in U_{b\mathcal{PT}}$, the eigenvalues emerge in complex
conjugate pairs.

From the above it is clear that we may define equally well different types
of $\mathcal{PT}$-operators closely related to the one introduced in (\ref%
{PPP}). For instance we can define 
\begin{equation}
\mathcal{P}_{x}:=\prod_{i=1}^{N}\sigma _{i}^{x}\qquad \text{and\qquad }%
\mathcal{P}_{y}:=\prod_{i=1}^{N}\sigma _{i}^{y},  \label{Pxy}
\end{equation}
which obviously act as 
\begin{equation}
\mathcal{P}_{x}:(\sigma _{i}^{x},\sigma _{i}^{y},\sigma _{i}^{z})\rightarrow
(\sigma _{i}^{x},-\sigma _{i}^{y},-\sigma _{i}^{z})\qquad \text{and\qquad }%
\mathcal{P}_{y}:(\sigma _{i}^{x},\sigma _{i}^{y},\sigma _{i}^{z})\rightarrow
(-\sigma _{i}^{x},\sigma _{i}^{y},-\sigma _{i}^{z}).
\end{equation}
Clearly these parity operators can not be used in the same way as $\mathcal{P%
}$ in (\ref{PPP}) to introduce a $\mathcal{PT}$-symmetry for $H(\lambda
,\kappa )$ when keeping $\mathcal{T}$ unchanged. However, they serve to
treat non-Hermitian Hamiltonians of a different kind, such as obvious
modifications of $H(\lambda ,\kappa )$ and also to allow for alternative
treatments of non-Hermitian spin chains, such as the XXZ-spin-chain in a
magnetic field \cite{Chico} 
\begin{equation}
H_{XXZ}=\frac{1}{2}\sum_{i=1}^{N-1}\left[ (\sigma _{i}^{x}\sigma
_{i+1}^{x}+\sigma _{i}^{y}\sigma _{i+1}^{y}+\Delta _{+}(\sigma
_{i}^{z}\sigma _{i+1}^{z}-1)\right] +\frac{\Delta _{-}}{2}(\sigma
_{1}^{z}-\sigma _{N}^{z}),  \label{XXZ}
\end{equation}
with $\Delta _{\pm }=(q\pm q^{-1})/2$ previously studied in \cite%
{Korff:2007qg,CKPT}. Obviously when $q\notin \mathbb{R}$ this Hamiltonian is
non-Hermitian, but we observe that it is $\mathcal{PT}$-symmetric when using
any of the parity operators defined in (\ref{Pxy}) and keeping $\mathcal{T}$
to be the usual complex conjugation 
\begin{equation}
\left[ \mathcal{P}_{x}\mathcal{T},H_{XXZ}\right] =0\qquad \text{and\qquad }%
\left[ \mathcal{P}_{y}\mathcal{T},H_{XXZ}\right] =0.  \label{sym}
\end{equation}
Thus besides reflecting the chain across its midpoint in form of a
\textquotedblleft macro-reflections\textquotedblright , as suggested in \cite%
{Korff:2007qg}, we may also carry out the parity transformations on each
individual side. It appears that these \textquotedblleft
micro-reflections\textquotedblright\ (\ref{PPP}), (\ref{Pxy}) allow for a
wider range of possibilities, such as for instance Hamiltonians of the type $%
H(\lambda ,\kappa )$ in (\ref{H}), which could not be tackled with $\mathcal{%
P}^{\prime }:\sigma _{i}^{x,y,z}\rightarrow \sigma _{N+1-i}^{x,y,z}$. The
different possibilities are simply manifestations of the well known
ambiguities non-Hermitian Hamiltonians possess with regard to their operator
content \cite{Urubu}. This also means that the symmetries (\ref{sym}) will
lead to a different kind of physical systems than those identified in \cite%
{Korff:2007qg}.

It is well known that $H_{XXZ}$ can be expressed in terms of generators of a
Temperley-Lieb algebra $E_{i}$, i.e. simply by writing the Hamiltonian
alternatively as $H_{XXZ}=$ $\sum_{i=1}^{N-1}E_{i}$. It is then trivial to
see that the algebra remains invariant under a $\mathcal{PT}$-transformation
when realized as (\ref{Pxy}): $\mathcal{T}:E_{i}\rightarrow E_{i}^{\ast }$, $%
\mathcal{P}_{x,y}:E_{i}\rightarrow E_{i}^{\ast }$, such that $\mathcal{P}%
_{x,y}\mathcal{T}:E_{i}\rightarrow E_{i}$. On the other hand when
implementing the \textquotedblleft macro-reflection\textquotedblright\ on
the entire chain, the $\mathcal{P}^{\prime }\mathcal{T}$-symmetry on the
generators is broken, i.e. $\mathcal{P}^{\prime }\mathcal{T}%
:E_{i}\rightarrow E_{N+1-i}$, as was found in \cite{Korff:2007qg}.

A further interesting non-Hermitian quantum spin chain has recently been
investigated by Deguchi and Ghosh \cite{DeGosh} 
\begin{equation}
H_{DG}=\sum_{i=1}^{N}\kappa _{zz}\sigma _{i}^{z}\sigma _{i+1}^{z}+\kappa
_{x}\sigma _{i}^{x}+\kappa _{y}\sigma _{i}^{y},  \label{DeGo}
\end{equation}%
with $\kappa _{zz}\in \mathbb{R}$ and $\kappa _{x}$, $\kappa _{y}\in \mathbb{%
C}$. Clearly when $\kappa _{x}$ or $\kappa _{y}\notin \mathbb{R}$ the
Hamiltonian $H_{DG}$ is not Hermitian, which is the case we will consider.
As the previous model also the quasi-Hermitian transverse Ising model allows
for different types of realizations for the $\mathcal{PT}$-symmetry. We
easily observe that the macro-reflections can not be implemented%
\begin{equation}
\left[ \mathcal{P}^{\prime }\mathcal{T},H\right] \neq 0,
\end{equation}%
whereas all the micro-reflections can be realized%
\begin{equation}
\left[ \mathcal{PT},H\right] =0\text{ \ \ for }\kappa _{x},\kappa _{y}\in i%
\mathbb{R},\quad \left[ \mathcal{P}_{x/y}\mathcal{T},H\right] =0\text{ \ \
for }\kappa _{x/y}\in \mathbb{R},\kappa _{y/x}\in i\mathbb{R}\text{.}
\end{equation}%
Once again these different possibilities raise the question about the unique
of the operator content in the model.

Having an explanation for the nature of the eigenvalue spectra, it is left
to show that one may in addition construct a meaningful metric for this
Hamiltonian with well defined quantum mechanical observables associated to
it. As already indicated, the metric is not even expected to be unique so
that, unlike as for the Hermitian case, the observables are no longer
defined by the Hamiltonian alone \cite{Urubu}. It remains therefore
ambiguous what Hamiltonians of the type $H(\lambda ,\kappa )$ describe in
terms of physical observables. Having constructed a metric one may often
also compute an isospectral Hermitian counterpart for $H(\lambda ,\kappa )$
for which the physical observables have the standard meaning.

One of the main purposes of this manuscript is that of finding the Hermitian
counterparts of the Hamiltonian (\ref{H}) and studying in some detail (at
least for small $N$) how many such Hermitian Hamiltonians can be constructed.

\section{Generalities}

\subsection{A new metric and an isospectral Hermitian partner from $\mathcal{%
PT}$-symmetry}

For the sake of self-consistency, we briefly recall the well known procedure 
\cite{EW,Bender:1998ke,Bender:2002vv,SW,Bender:2004sa} of how to construct a
meaningful metric and isospectral Hermitian counterpart, $h$, for a
non-Hermitian Hamiltonian, $H$. We assume the Hamiltonian to be
diagonalizable and to possess a discrete spectrum. Being non-Hermitian the
Hamiltonian has non identical left $\left| \Phi \right\rangle $ and right
eigenvectors $\left| \Psi \right\rangle $ with eigenvalue equations 
\begin{equation}
H\left| \Phi _{n}\right\rangle =\varepsilon _{n}\left| \Phi
_{n}\right\rangle \qquad \text{and\qquad }H^{\dagger }\left| \Psi
_{n}\right\rangle =\epsilon _{n}\left| \Psi _{n}\right\rangle \text{\qquad
for }n\in \mathbb{N}.
\end{equation}
The eigenvectors are in general not orthogonal $\left\langle \Phi
_{n}\right. \left| \Phi _{m}\right\rangle \neq \delta _{nm}$, but form a
biorthonormal basis 
\begin{equation}
\left\langle \Psi _{n}\right. \left| \Phi _{m}\right\rangle =\delta
_{nm},\qquad \sum_{n}\left| \Psi _{n}\right\rangle \left\langle \Phi
_{n}\right| =\mathbb{I}.
\end{equation}
We assume the existence of a selfadjoint, but not necessarily positive,
parity operator $\mathcal{P}$ whose adjoint action conjugates the
Hamiltonian 
\begin{equation}
H^{\dagger }=\mathcal{P}H\mathcal{P}\qquad \text{with}\qquad \mathcal{P}^{2}=%
\mathbb{I}.  \label{P}
\end{equation}
The action of this operator on the eigenvectors 
\begin{equation}
\mathcal{P}\left| \Phi _{n}\right\rangle =s_{n}\left| \Psi _{n}\right\rangle 
\text{\qquad with }s_{n}=\pm 1  \label{S}
\end{equation}
defines the signature $s=(s_{1},s_{2},\ldots ,s_{n})$, which serves to
introduce the so-called $\mathcal{C}$-operator\footnote{%
The is an unfortunate notation and it should be pointed out that the
operator is not related to the standard charge conjugation operator in
quantum field theory.} 
\begin{equation}
\mathcal{C}:=\sum_{n}s_{n}\left| \Phi _{n}\right\rangle \left\langle \Psi
_{n}\right| ,  \label{C}
\end{equation}
satisfying 
\begin{equation}
\left[ \mathcal{C},H\right] =0,\qquad \left[ \mathcal{C},\mathcal{PT}\right]
=0,\qquad \mathcal{C}^{2}=\mathbb{I}.  \label{CP}
\end{equation}
Next we employ this operator to define a new operator $\rho $, which also
relates the Hamiltonian to its conjugate 
\begin{equation}
\rho :=\mathcal{PC},\mathcal{\qquad }H^{\dagger }\rho =\rho H.  \label{rel}
\end{equation}
Depending now on the assumptions made for $\rho $, such systems allow for
different types of conclusions. When $\rho $ is positive and Hermitian, but
not necessarily invertible, the system is referred to as quasi-Hermitian 
\cite{Dieu,Urubu}. In this case the existence of a definite metric is
guaranteed and the eigenvalues are real. In turn when $\rho $ is invertible
and Hermitian, but not necessarily positive, the system is called
pseudo-Hermitian \cite{pseudo1,pseudo2,Mostafazadeh:2001nr}. For this type
of scenario the eigenvalues are always real but no definite conclusions can
be made with regard to the existence of a definite metric. Here we will
identify operators $\rho $ which are quasi-Hermitian as well as
pseudo-Hermitian.

Finally we may factorize $\rho $ into a new operator\footnote{%
When $\eta $ is Hermitian, it just corresponds to a Dyson transformation 
\cite{Dyson} employed in the so-called Holstein-Primakov method \cite%
{Holstein}. For practical purposes it is useful to have a name for this
operator and therefore we refer to $\eta $ from now on as the Dyson map.} $%
\eta $ and use it to construct an isospectral Hermitian counterpart for $H$ 
\begin{equation}
h=\eta H\eta ^{-1}=h^{\dagger }\text{ \qquad }\Leftrightarrow \qquad
H^{\dagger }=\rho H\rho ^{-1}\quad \text{with }\rho =\eta ^{\dagger }\eta 
\text{.}  \label{1}
\end{equation}%
In other words assuming the existence of an inverse for $\rho $ and its
factorization in form of (\ref{1}) one can derive a Hermitian counterpart $h$
for $H$ and vice versa.

\subsection{Expectation values of local observables \label{local}}

\noindent As discussed above, when dealing with non-Hermitian Hamiltonians
the standard metric is generally indefinite and therefore a new, physically
sensible, metric needs to be defined by means of the construction described
before. This amounts to introducing a new inner product $\langle \quad
|\quad \rangle _{\rho }$ which is defined in terms of the standard inner
product $\langle \quad |\quad \rangle $ as 
\begin{equation}
\langle \Phi |\Psi \rangle _{\rho }:=\langle \Phi |\rho \Psi \rangle ,
\end{equation}
for arbitrary states, $\langle \Phi |$ and $|\Psi \rangle $. Assuming that
all local operators $\mathcal{O}$ in the non-Hermitian theory are related to
their counterparts $o$ in the Hermitian theory in the same manner as the
corresponding Hamiltonians 
\begin{equation}
\eta \mathcal{O}\eta ^{-1}=o,
\end{equation}
one finds that a generic matrix element of the operator $\mathcal{O}$ has
the form, 
\begin{equation}
\langle \Phi |\rho \mathcal{O}|\Psi \rangle =\langle \Phi |\eta ^{\dagger
}o\eta |\Psi \rangle =\langle {\phi }|o|\psi \rangle ,  \label{ff}
\end{equation}
where $|\Psi \rangle $ and $\langle \Phi |$ are eigenstates of the
non-Hermitian Hamiltonian and its conjugate, respectively. The states $|\psi
\rangle $ and $\langle \phi |$ are related to the previous two states by $%
|\psi \rangle =\eta |\Psi \rangle $ and $\langle \phi |=\langle \Phi |\eta
^{\dagger }$, that is, they are eigenstates of the Hermitian Hamiltonian
corresponding to the same eigenvalues. Equation (\ref{ff}) will be used
later on in this paper for the computation of various kinds of expectation
values.

\subsection{Perturbation theory \label{pert}}

In most cases the above mentioned operators can not be computed exactly and
one has to resort to a perturbative analysis. Let us recall the main
features of such a treatment. To start with it is convenient to separate the
Hamiltonian into its Hermitian and non-Hermitian part as $H(\lambda ,\kappa
)=h_{0}(\lambda )+i\kappa h_{1}$, where $h_{0}$ and $h_{1}$ are both
Hermitian with $\kappa $ being a real coupling constant. The latter term may
then be treated as the perturbing term. For our concrete case (\ref{H}) the
individual components are 
\begin{equation}
h_{0}(\lambda )=-\frac{1}{2}\sum_{i=1}^{N}(\sigma _{i}^{z}+\lambda \sigma
_{i}^{x}\sigma _{i+1}^{x}),\qquad \text{and}\qquad h_{1}=-\frac{1}{2}%
\sum_{i=1}^{N}\sigma _{i}^{x},  \label{h0h1}
\end{equation}
such that $h_{0}(\lambda )$ corresponds to the Ising spin chain coupled to a
magnetic field in the $z$ direction and the perturbing term is an imaginary
magnetic field in the $x$-direction. In order to determine $\eta $, $\rho $
and $h$ we can now solve either of the two equations in (\ref{1}). Here we
decide to commence with the latter. Making the further assumption that $\eta 
$ is Hermitian and of the form $\eta =e^{q/2}$ this amounts to solving 
\begin{equation}
H^{\dagger }=e^{q}He^{-q}=H+[q,H]+\frac{1}{2}[q,[q,H]]+\frac{1}{3!}%
[q,[q,[q,H]]]+\cdots  \label{per}
\end{equation}
where we have employed the Backer-Campbell-Hausdorff identity. Writing $H$
and $H^{\dagger }$ in terms of $h_{0}$ and $h_{1}$ equation (\ref{per})
becomes 
\begin{equation}
2i\kappa h_{1}+i\kappa \lbrack q,h_{1}]+\frac{i\kappa }{2}%
[q,[q,h_{1}]]+\cdots =[h_{0},q]+\frac{1}{2}[q,[h_{0},q]]+\cdots
\label{perh01}
\end{equation}
For most non-Hermitian Hamiltonians, such as for our model (\ref{H}), this
equation is very difficult to solve for $q$. When the $(\ell +1)$-fold
commutator of $q$ with $h_{0}$, denoted by $c_{q}^{(\ell +1)}(h_{0})$
vanishes, \ closed formulae were found in \cite{CA} 
\begin{equation}
h=h_{0}+\sum\limits_{n=1}^{[\frac{\ell }{2}]}\frac{(-1)^{n}E_{n}}{4^{n}(2n)!}%
c_{q}^{(2n)}(h_{0}),\quad H=h_{0}-\sum\limits_{n=1}^{[\frac{\ell +1}{2}]}%
\frac{\kappa _{2n-1}}{(2n-1)!}c_{q}^{(2n-1)}(h_{0}),  \label{HH}
\end{equation}
where $\left[ x\right] $ denotes the integer part of a number $x$ and $E_{n}$
are Euler's numbers, e.g. $E_{1}=1$, $E_{2}=5$, $E_{3}=61$, $%
E_{4}=1385,\ldots $ The coefficients $\kappa _{2n-1}$ were determined by
means of a recursive equation, which was solved by 
\begin{equation}
\kappa _{n}=\frac{1}{2^{n}}\sum\limits_{m=1}^{\left[ (n+1)/2\right]
}(-1)^{n+m}\binom{n}{2m}E_{m},  \label{kan}
\end{equation}
such that $\kappa _{1}=1/2,\kappa _{3}=-1/4,\kappa _{5}=1/2,\kappa
_{7}=-17/8,\ldots $

One may also impose some further structure on $q$ and expand it as 
\begin{equation}
q=\sum_{k=1}^{\infty }\kappa ^{2k-1}q_{2k-1},  \label{eta12}
\end{equation}
so that each perturbative contribution $q_{2k-1}$ is a $\kappa $-independent
matrix. For models of the form considered here only odd powers of $\kappa $
appear in the perturbative expansion. This is essentially due to the fact
that $H$ and $H^{\dagger }$ are related to each other by $\kappa \rightarrow
-\kappa $. Substituting the expansion (\ref{eta12}) into the equation (\ref%
{perh01}) one finds a set of equations for $q_{1},q_{3},q_{5},\ldots $ by
equating those terms in (\ref{perh01}) which are of the same order in
perturbation theory in $\kappa $. The first few equations are given by 
\begin{eqnarray}
&&[h_{0},q_{1}]=2ih_{1},  \label{c1} \\
&&[h_{0},q_{3}]=\frac{i}{6}[q_{1},[q_{1},h_{1}]],  \label{c3} \\
&&[h_{0},q_{5}]=\frac{i}{6}[q_{1},[q_{3},h_{1}]]+\frac{i}{6}%
[q_{3},[q_{1},h_{1}]]-\frac{i}{360}[q_{1},[q_{1},[q_{1},[q_{1},h_{1}]]]].
\label{c5}
\end{eqnarray}
As we can see easily, they can be solved recursively, namely once $q_{1}$ is
known, one case solve for $q_{3}$ and so on. A closed expression for the
commutator $[h_{0},q_{n}]$ in terms of commutators $[q_{m},h_{1}]$ with $m<n$
was derived in \cite{CKPT}. Perturbation theory has been carried out in the
past for various non-Hermitian models, e.g. \cite%
{Bender:2004sa,CA,Mosta,ACIso,Can,CKPT}.

The model at hand is special in the sense that it involves two coupling
constants, i.e. $\kappa $ and $\lambda $, such that it allows for an
alternative perturbative expansion in terms of the latter. Indeed we will
demonstrate below that the case $\lambda =0$ can be solved exactly and we
can therefore expand around that solution. Proceeding similarly as for the $%
\kappa $-perturbation theory we separate the Hamiltonian into its single
spin contribution and into the nearest neighbour interaction term $H(\lambda
,\kappa )=\tilde{H}_{0}(\kappa )+\lambda \tilde{h}_{1}$with 
\begin{equation}
\tilde{H}_{0}(\kappa )=-\frac{1}{2}\sum_{i=1}^{N}(\sigma _{i}^{z}+i\kappa
\sigma _{i}^{x})\quad \text{and\quad }\tilde{h}_{1}=-\frac{1}{2}%
\sum_{i=1}^{N}\sigma _{i}^{x}\sigma _{i+1}^{x}.  \label{Hh}
\end{equation}%
We stress that the counterparts of (\ref{c1})-(\ref{c5}) in the well known $%
\kappa $-expansion explained above differ substantially in the $\lambda $%
-expansion. The details will be explained in the main part of the manuscript
below. Having the option to construct two perturbative series, we in
principle have in addition the possibility to combine them in a manner that
has proved to be very successful in the context of high intensity laser
physics \cite{AC1}.

\subsection{Ambiguities in the physical observables}

\label{ambi}

As mentioned previously, one can argue that the metric $\rho $ is not
unique. In the perturbation theory framework, this can be easily seen from
the fact that the equations (\ref{c1})-(\ref{c5}) (and any other equations
arising at higher orders in perturbation theory) admit many different
solutions. The non-uniqueness of $\eta $ or, equivalently, the fact that
several independent Hermitian Hamiltonians $h$ may exist which are all
related to the same non-Hermitian Hamiltonian by different unitary
transformations is well known in the literature. Indeed, this fact has been
noticed already in the past \cite{Urubu,MGH,ACIso,Mostsyme,PEGAAF,PEGAAF2}
and is currently still object of debate \cite{bendernew,kleefeld}.

Assuming now the Dyson map $\eta $ in (\ref{1}) to be Hermitian and related
to the operators $\mathcal{P}$, $\mathcal{C}$ and $\rho $ as defined in (\ref%
{rel}) we simply obtain 
\begin{equation}
\eta =\eta ^{\dagger }\qquad \Rightarrow \qquad \eta ^{2}=\rho =\mathcal{PC}.
\label{etahermi}
\end{equation}%
Writing $\eta =e^{q/2}$, it is obvious that we can always add to $q$ any
matrix $b$ that commutes with the full Hamiltonian $[H,b]=0$ and with $q$, $%
[q,b]=0$ 
\begin{equation}
h=e^{q/2+b}He^{-q/2-b}=e^{q/2}He^{-q/2},  \label{deg}
\end{equation}%
and still solve equations (\ref{1}). This kind of ambiguity is not very
interesting, as it will not change $h$ and therefore not lead to new
physics. A somewhat less trivial ambiguity was pointed out in \cite{ACIso},
which will generate different types of Hermitian counter-parts to $H$. It
originates from the fact that we can always add to $q_{1},q_{3},q_{5},\ldots 
$ any matrix commuting with $h_{0}$ as we may easily observe in equations (%
\ref{c1})-(\ref{c5}). Below we will see that in principle for specific
examples many such matrices can be found.

However, by relating $\eta $ to the operators $\mathcal{C}$ and $\mathcal{P}$
as in (\ref{etahermi}) we are introducing further constraints on the form of 
$\eta $. These constraints follow from the equations (\ref{CP}),
particularly the last two equations there. Using the explicit form (\ref%
{etahermi}) they can be rewritten as 
\begin{equation}
\mathcal{PT}e^{q}\mathcal{PT}=e^{q},\qquad \mathcal{P}e^{q}\mathcal{P}%
=e^{-q}.  \label{ecu}
\end{equation}
by employing the equality $\mathcal{C}=\eta ^{2}\mathcal{P}=e^{q}\mathcal{P}$%
. \ In order for (\ref{ecu}) to be satisfied, it is required that 
\begin{equation}
\mathcal{P}q\mathcal{P}=\mathcal{T}q\mathcal{T}=-q,  \label{cons1}
\end{equation}
and consequently 
\begin{equation}
\mathcal{P}q_{2k-1}\mathcal{P}=\mathcal{T}q_{2k-1}\mathcal{T}%
=-q_{2k-1},\qquad \forall \quad k\in \mathbb{Z}^{+}.  \label{cons}
\end{equation}
Below, we will show that these constraints are sufficient in many cases to
fix the operator $\eta $ and therefore the metric completely. However, it
should be noted that these arguments are based on the assumption that $\rho $
acquires the form (\ref{etahermi}) and furthermore that the parity operator
is unique, which as we exemplified (\ref{Pxy}) is not always the case.

\section{The Yang-Lee quantum chain: non perturbative results\label{exact}}

\label{4}

We will now employ the general ideas and definitions introduced in the
previous subsection for the quantum spin chain Hamiltonian (\ref{H}). In
particular, we will show how to obtain exact solutions for the operators $%
\eta $, $\rho $ and $h$ in the two particular situations: i) $\lambda $ or $%
\kappa $ are vanishing and $N$ is generic and ii) $\lambda $ and $\kappa $
are arbitrary and $N$ is taken to be small.

For large values of $N$ it will be convenient to use the following
abbreviation 
\begin{equation}
S_{a_{1}a_{2}\ldots a_{p}}^{N}:=\sum_{k=1}^{N}\sigma _{k}^{a_{1}}\sigma
_{k+1}^{a_{2}}\ldots \sigma _{k+p-1}^{a_{p}},\qquad \text{for}\quad
a_{i}=x,y,z,u;\quad i=1,\ldots ,p\leq N.  \label{matrices}
\end{equation}%
We denote here $\sigma ^{u}=\mathbb{I}$ to allow for non-local, i.e. not
nearest neighbour, interactions. In this notation the Hamiltonian (\ref{H})
reads 
\begin{equation}
H(\lambda ,\kappa )=h_{0}(\lambda )+i\kappa h_{1},\quad \text{with}\quad
h_{0}(\lambda )=-\frac{1}{2}(S_{z}^{N}+\lambda S_{xx}^{N}),\qquad h_{1}=-%
\frac{1}{2}S_{x}^{N}.
\end{equation}%
In what follows it will also be important to use the adjoint action of $%
\mathcal{P}$, $\mathcal{T}$ and $\mathcal{PT}$ on the generators $%
S_{a_{1}a_{2}\ldots a_{p}}^{N}$. It is easy to compute 
\begin{eqnarray}
\mathcal{P}S_{a_{1}a_{2}\ldots a_{p}}^{N}\mathcal{P}
&=&(-1)^{n_{y}+n_{x}}S_{a_{1}a_{2}\ldots a_{p}}^{N},  \label{pmat} \\
\mathcal{T}S_{a_{1}a_{2}\ldots a_{p}}^{N}\mathcal{T}
&=&(-1)^{n_{y}}S_{a_{1}a_{2}\ldots a_{p}}^{N},  \label{tmat} \\
\mathcal{PT}S_{a_{1}a_{2}\ldots a_{p}}^{N}\mathcal{PT}
&=&(-1)^{n_{x}}S_{a_{1}a_{2}\ldots a_{p}}^{N},  \label{ptmat}
\end{eqnarray}%
where $n_{x}$, $n_{y}$ are the numbers of indices $a_{i}$ equal to $x$, $y$,
respectively. These identities follow directly from the definitions (\ref%
{eff}) and (\ref{teff}).

\subsection{Limiting cases: $\protect\lambda =0$ or $\protect\kappa =0$\label%
{known}}

\label{limit}

\noindent Let us start by considering the special case $\lambda =0$ for
which 
\begin{equation}
h_{0}(0)=-\frac{1}{2}S_{z}^{N}\quad \text{and}\quad h_{1}=-\frac{1}{2}%
S_{x}^{N}.
\end{equation}%
Although the Hamiltonian is extremely simple, it is still non-Hermitian, and
thus serves as a benchmark to illustrate the above mentioned notions. For
example, a matrix $\eta $ that relates $H(0,\kappa )$ to its Hermitian
counterpart $h(0,\kappa )$ is easily found to be 
\begin{equation}
\eta =e^{q/2}=e^{-\frac{1}{2}\text{arctanh}(\kappa )S_{y}^{N}}.  \label{w0}
\end{equation}%
Its adjoint action on $S_{x}^{N}$ and $S_{z}^{N}$ is simply 
\begin{equation}
\eta S_{x}^{N}\eta ^{-1}=\frac{1}{\sqrt{1-\kappa ^{2}}}(i\kappa
S_{z}^{N}+S_{x}^{N}),\qquad \eta S_{z}^{N}\eta ^{-1}=\frac{1}{\sqrt{1-\kappa
^{2}}}(S_{z}^{N}-i\kappa S_{x}^{N}),  \label{acts}
\end{equation}%
which when we evaluate (\ref{1}) yields the Hermitian counterpart to $\tilde{%
H}_{0}(\kappa )$ in (\ref{Hh}) 
\begin{equation}
h(0,\kappa )=-\frac{1}{2}\sqrt{1-\kappa ^{2}}S_{z}^{N}.  \label{final}
\end{equation}%
This Hamiltonian describes a spin chain for which no mutual interaction
between spins along the chain occurs. An external magnetic field is applied
at each site of the chain, whose intensity is governed by the value of $%
\kappa $ and is the same at every site. The constraint $-1<\kappa <1$
ensures the Hamiltonian $h(0,\kappa )$ and $\eta $ to be Hermitian. Given
the simplicity of $h(0,\kappa )$ we can easily find its full set of
eigenstates and eigenvalues, hence those of $H(0,\kappa )$. The operator $%
S_{z}^{N}$ is a diagonal matrix with entries 
\begin{equation}
S_{z}^{N}=\text{diag}(N,N-2,\ldots ,-N+2,-N).  \label{diag}
\end{equation}%
The entries in the diagonal (eigenvalues) are $N-2p$ with $p=0,\ldots ,N$.
They are not necessarily in decreasing order and, except for $N$ and $-N$,
all other eigenvalues are degenerate. For example, the eigenvalues $N-2$ and 
$2-N$ are always $N$ times degenerate. This means that there is a single
ground state with minimum energy, 
\begin{equation}
E_{g}(\kappa )=-\frac{N}{2}\sqrt{1-\kappa ^{2}},  \label{eg}
\end{equation}%
and the corresponding eigenstate is simply 
\begin{equation}
|\psi _{g}\rangle =\bigotimes_{i=1}^{N}\left( 
\begin{array}{c}
1 \\ 
0%
\end{array}%
\right) _{i},  \label{gss}
\end{equation}%
associated to a configuration with all spins \textquotedblleft
up\textquotedblright , hence aligned with the magnetic field that is being
applied at each site of the chain.

The situation when $\kappa =0$ and $\lambda $ is arbitrary corresponds to
the Hermitian Hamiltonian given by $h_{0}(\lambda )$, that is the Ising spin
chain with a magnetic field in the $z$-direction. In this case, $\eta =%
\mathbb{I}$, which is automatically ensured when using perturbation theory.
The eigenstates and eigenvalues of this Hamiltonian have been studied in the
literature by using the Bethe ansatz approach, see e.g. \cite{baxter,izergin}%
. In particular, the ground state can not be written in such as simple form
as (\ref{gss}), as it will depend on the value of $\lambda $. One does know
however, that, for finite $N$, it will interpolate between the $\lambda =0$
case, in which the ground state is (\ref{gss}) and the $\lambda \rightarrow
\infty $ case, in which the ground state will correspond to alternating
up-down spins.

\subsubsection{Uniqueness of the Dyson operator}

\label{unih00} In light of the discussion in section \ref{ambi} it is also
interesting to investigate the uniqueness of (\ref{w0}). Indeed, we will now
show that (\ref{w0}) is the only solution to (\ref{1}) which is consistent
with (\ref{ecu}) for the Hamiltonian $H(0,\kappa )$. This can be proven in
two steps: firstly we will characterize the subset of matrices and linear
combinations thereof that satisfy (\ref{cons1}) and secondly, we will show
that none of these matrices can be in the kernel of $h_0(0)$. Let us define
the matrices, which provide a basis for the set of $2^N \otimes 2^N$%
-Hermitian matrices, 
\begin{equation}  \label{M}
M_{a_1\ldots a_N}=\sigma_1^{a_1}\otimes \cdots \otimes \sigma_N^{a_N},\quad 
\text{with} \quad a_i=x,y,z, \,\, \text{or} \,\, u \quad
\forall\,\,i=1,\ldots,N.
\end{equation}
Recall the definition $\sigma_i^u = \mathbb{I}_i$. Let us consider an
arbitrary linear combination of the matrices (\ref{M}). The action of parity
and time reversal on such a linear combination is analogous to (\ref{pmat})
and (\ref{tmat}). From this it follows that, in order for any linear
combination of matrices $M_{a_1,\ldots,a_p}$ to transform as $q$ does in
equations (\ref{cons1}) it must be such that for all matrices in the linear
combination $n_y$ is odd and $n_x$ is even ($n_x$ and $n_y$ as defined after
equation (\ref{ptmat})).

We will now argue that no matrix in the kernel of $h_0(0)$ is of this form.
There are various ways of having a vanishing commutator $[h_0(0),B]=0$. The
most obvious solution is for $B$ to be a diagonal matrix, as $h_0(0)$ is
itself diagonal. In terms of the matrices (\ref{M}), this means selecting
out those that are tensor products of $\sigma^z$ and $\mathbb{I}$ only.
There are overall $2^N$ such matrices and obviously none of them has $n_y$
odd. This would be sufficient to conclude that the solution (\ref{w0}) is
unique if only the kernel of $h_0(0)$ had dimension $2^N$. This is not so
because $h_0(0)$ has degenerate eigenvalues.

Any additional matrices in the kernel will be some linear combination of
matrices (\ref{M}) involving at least one index $x$ or $y$. Employing the
commutation relations (\ref{com}), it is easy to see that there are
basically two kinds of additional matrices that are in the kernel of $h_0(0)$%
: firstly, the matrices $M_{xyu\ldots u}-M_{yxu\ldots u}$ and
generalizations thereof , which are antisymmetric under the exchange of
indices $x\leftrightarrow y$ and violate the condition $n_x$ even and
secondly, the matrices $M_{xxu\ldots u}+M_{yy u \ldots u}$ and
generalizations thereof, which violate the condition $n_y$ odd and are
symmetric under the exchange of indices $x \leftrightarrow y$.
Generalizations of these matrices are those obtained by replacing any number
of indices $u$ by $z$ and/or permuting indices, as well as other matrices of
similar characteristics, such as $M_{xxxxu \ldots u}+M_{yyyyu \ldots
u}+M_{xxyyu \ldots u}+M_{yyxxu\ldots u}$ and so on. Since this is more an
argument than a proof, we would like to support it with two examples. For $%
N=2$ 
\begin{equation}  \label{h002}
h_0(0)=\text{diag}(-1,0,0,1),
\end{equation}
and the kernel has dimensions 6, as one eigenvalue is twice degenerate. It
is generated by the matrices 
\begin{equation}
M_{xy}-M_{yx},\quad M_{xx}+M_{yy}, \quad M_{zz},\quad M_{zu},\quad M_{uz}
\quad \text{and} \quad M_{uu}=\mathbb{I}.
\end{equation}
For $N=3$ we have that $h_0(0)$ has four different eigenvalues, two of which
are three times degenerate, 
\begin{equation}  \label{h003}
h_0(0)=\text{diag}\left(-\frac{3}{2},-\frac{1}{2},-\frac{1}{2},\frac{1}{2},-%
\frac{1}{2},\frac{1}{2},\frac{1}{2},\frac{3}{2}\right),
\end{equation}
The dimension of the kernel then becomes 20. Its generators are the matrices 
\begin{eqnarray}
&& M_{xyu}-M_{yxu},\quad M_{xuy}-M_{yux}, \quad M_{uxy}-M_{uyx},  \notag \\
&& M_{xxu}+M_{yyu},\quad M_{xux}+M_{yuy}, \quad M_{uxx}+M_{uyy},  \notag \\
&& M_{xyz}-M_{yxz},\quad M_{xzy}-M_{yzx}, \quad M_{zxy}-M_{zyx},  \notag \\
&& M_{xxz}+M_{yyz},\quad M_{xzx}+M_{yzy}, \quad M_{zxx}+M_{zyy},  \notag \\
&& M_{zzz}, \quad M_{zzu},\quad M_{zuz}, \quad M_{uzz}, \quad M_{zuu}, \quad
M_{uzu}, \quad M_{uuz},\quad M_{uuu},
\end{eqnarray}
As shown before, these examples confirm once more that no element in the
kernel of $h_0(0)$ can fulfill the conditions (\ref{cons1}) and therefore
could not be added to $q$, whilst fulfilling such conditions. Thus no
matrices in the kernel of $h_0(0)$ satisfy the conditions (\ref{cons1}) and
the solution (\ref{w0}) is unique if the operator $\eta=e^{q/2}$ is to be
Hermitian.

\subsection{The $N=2$ case: two sites}

\noindent We have already identified the $\mathcal{PT}$-symmetry for the
Hamiltonian (\ref{H}) with $\mathcal{P}$ given as specified in (\ref{PPP})
satisfying (\ref{P}). Let us now take the length of the spin chain to be $%
N=2 $ and compute the quantities as outlined in the previous section.

For two sites we may chose without loss of generality the boundary
conditions to be periodic $\sigma _{N+1}^{x}=\sigma _{1}^{x}$ as any other
choice may be achieved simply by a re-definition of $\lambda $. In this case
the Hamiltonian (\ref{H}) acquires the simple form of a non-Hermitian $%
4\times 4$-matrix. In order to make notations clear, we will write this
matrix here in the various notations introduced so far, 
\begin{eqnarray}
H(\lambda ,\kappa ) &=&-\frac{1}{2}\left[ \sigma ^{z}\otimes \mathbb{I}+%
\mathbb{I}\otimes \sigma ^{z}+2\lambda \sigma ^{x}\otimes \sigma
^{x}+i\kappa \left( \mathbb{I}\otimes \sigma ^{x}+\sigma ^{x}\otimes \mathbb{%
I}\right) \right] ,  \notag \\
&=&-\frac{1}{2}\left[ \sigma _{1}^{z}+\sigma _{2}^{z}+2\lambda \sigma
_{1}^{x}\sigma _{2}^{x}+i\kappa \left( \sigma _{2}^{x}+\sigma
_{1}^{x}\right) \right] ,  \notag \\
&=&-\frac{1}{2}[S_{z}^{2}+\lambda S_{xx}^{2}+i\kappa S_{z}^{2}]=-\left( 
\begin{array}{rrrr}
-1 & \frac{i\kappa }{2} & \frac{i\kappa }{2} & \lambda \\ 
\frac{i\kappa }{2} & 0 & \lambda & \frac{i\kappa }{2} \\ 
\frac{i\kappa }{2} & \lambda & 0 & \frac{i\kappa }{2} \\ 
\lambda & \frac{i\kappa }{2} & \frac{i\kappa }{2} & -1%
\end{array}%
\right) ,  \label{H22}
\end{eqnarray}%
where the first line shows the most explicit way of writing the Hamiltonian,
the second line shows a simplified version, were the tensor products are
omitted and absorbed into the $\sigma $s as specified after (\ref{H}). The
last line uses the notation introduced in (\ref{matrices}).

\begin{center}
\includegraphics[width=9cm,height=6cm]{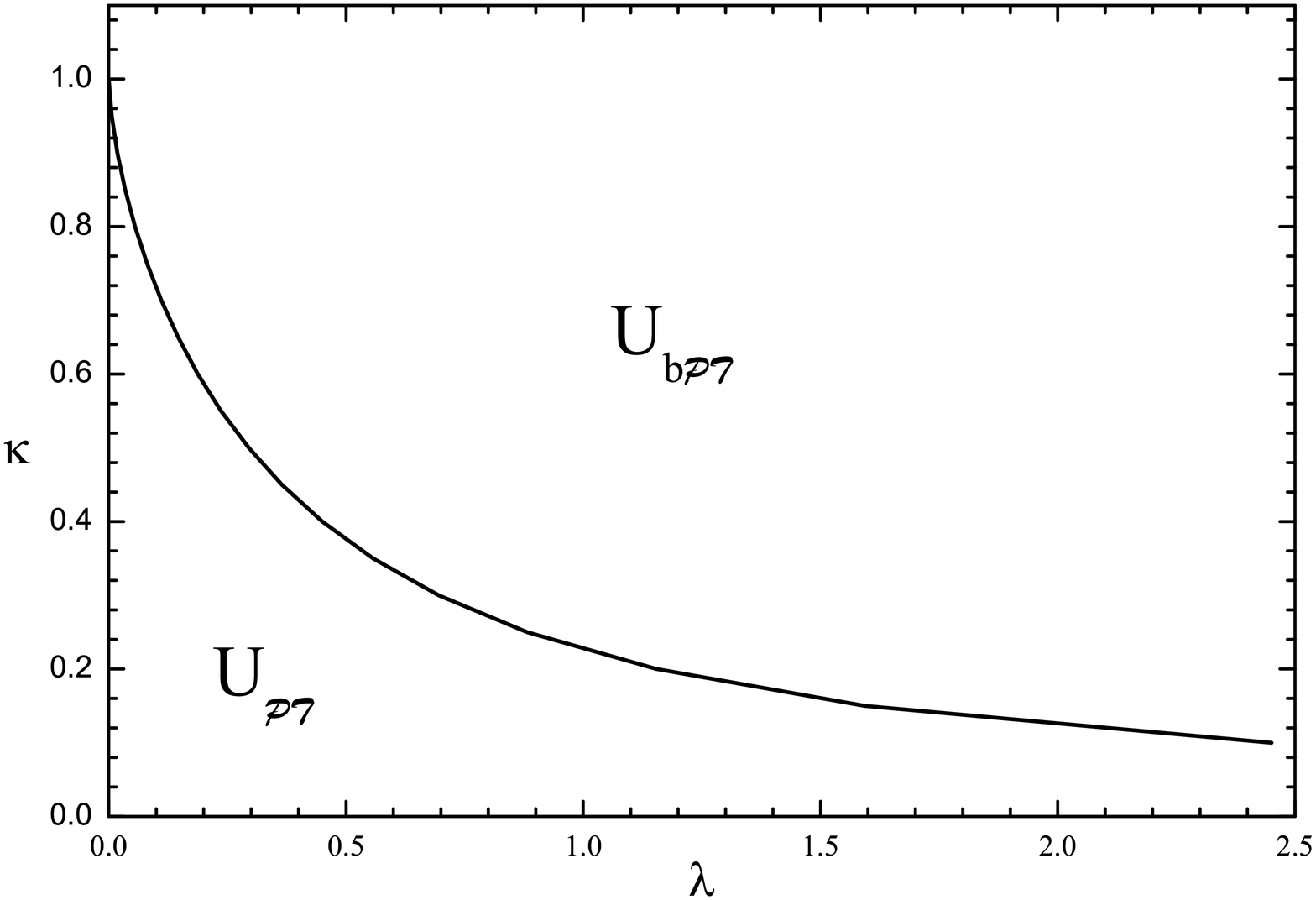}
\end{center}

\noindent {\small {\textbf{Figure 1:} Domains of broken and unbroken $%
\mathcal{PT}$-symmetry}}

At first we shall be concerned with the spectral properties of this
Hamiltonian. The two subdomains $U_{\mathcal{PT}}$ and $U_{b\mathcal{PT}}$
,\ as introduced in (\ref{U}), have already been identified numerically in 
\cite{gehlen1} for spin chain lengths up to $N=19$, that is for matrices up
to the remarkable size of $524288\times 524288$. For $N=2$ the eigenvalues
for (\ref{H22}) are easily computed analytically as the characteristic
polynomial factorizes into a third and first order polynomial. The
discriminant\ $\Delta $ of the third order polynomial is computed by 
\begin{equation}
\Delta =r^{2}-q^{3}\text{\qquad with \ }q=\frac{1}{9}\left( -3\kappa
^{2}+4\lambda ^{2}+3\right) ,\quad r=\frac{\lambda }{27}\left( 18\kappa
^{2}+8\lambda ^{2}+9\right) .
\end{equation}%
The eigenvalues are guaranteed to be real when the discriminant is smaller
or equal to zero, such that $U_{\mathcal{PT}}$ \ is defined as 
\begin{equation}
U_{\mathcal{PT}}=\left\{ \lambda ,\kappa :\Delta =\kappa ^{6}+8\lambda
^{2}\kappa ^{4}-3\kappa ^{4}+16\lambda ^{4}\kappa ^{2}+20\lambda ^{2}\kappa
^{2}+3\kappa ^{2}-\lambda ^{2}-1\leq 0\right\} .  \label{upt}
\end{equation}%
The regions $U_{\mathcal{PT}}$ and $U_{b\mathcal{PT}}$ are depicted in
Figure 1, from which we note that in order to have a real eigenvalue
spectrum $\kappa $ is restricted to take values between $0$ and $1$, whereas 
$\lambda $ is left unbounded $\lambda \in \lbrack 0,\infty )$.

\noindent The four real eigenvalues are then computed to 
\begin{equation}
\begin{array}{ccc}
\varepsilon _{1}=\lambda ,\qquad & \varepsilon _{2}=2q^{\frac{1}{2}}\cos
\left( \frac{\theta }{3}\right) -\frac{\lambda }{3},\qquad & \varepsilon
_{3,4}=2q^{\frac{1}{2}}\cos \left( \frac{\theta }{3}+\pi \mp \frac{2\pi }{3}%
\right) -\frac{\lambda }{3},%
\end{array}%
\end{equation}
where the additional abbreviation $\theta =\arccos \left( r/q^{3/2}\right) $
has been introduced. We depict these eigenvalues in Figure 2,

\noindent \includegraphics[width=7.5cm,height=6cm]{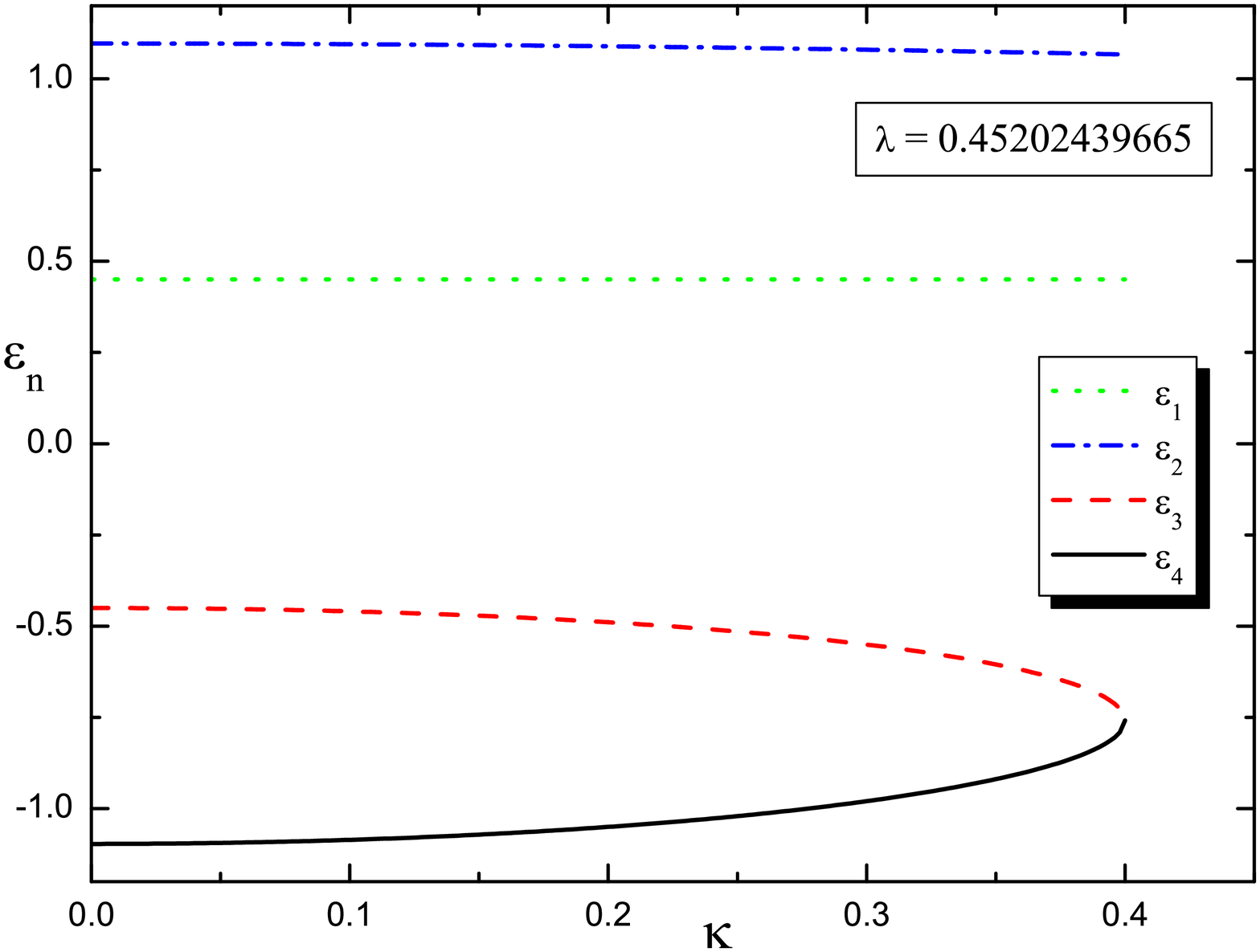} %
\includegraphics[width=7.5cm,height=6cm]{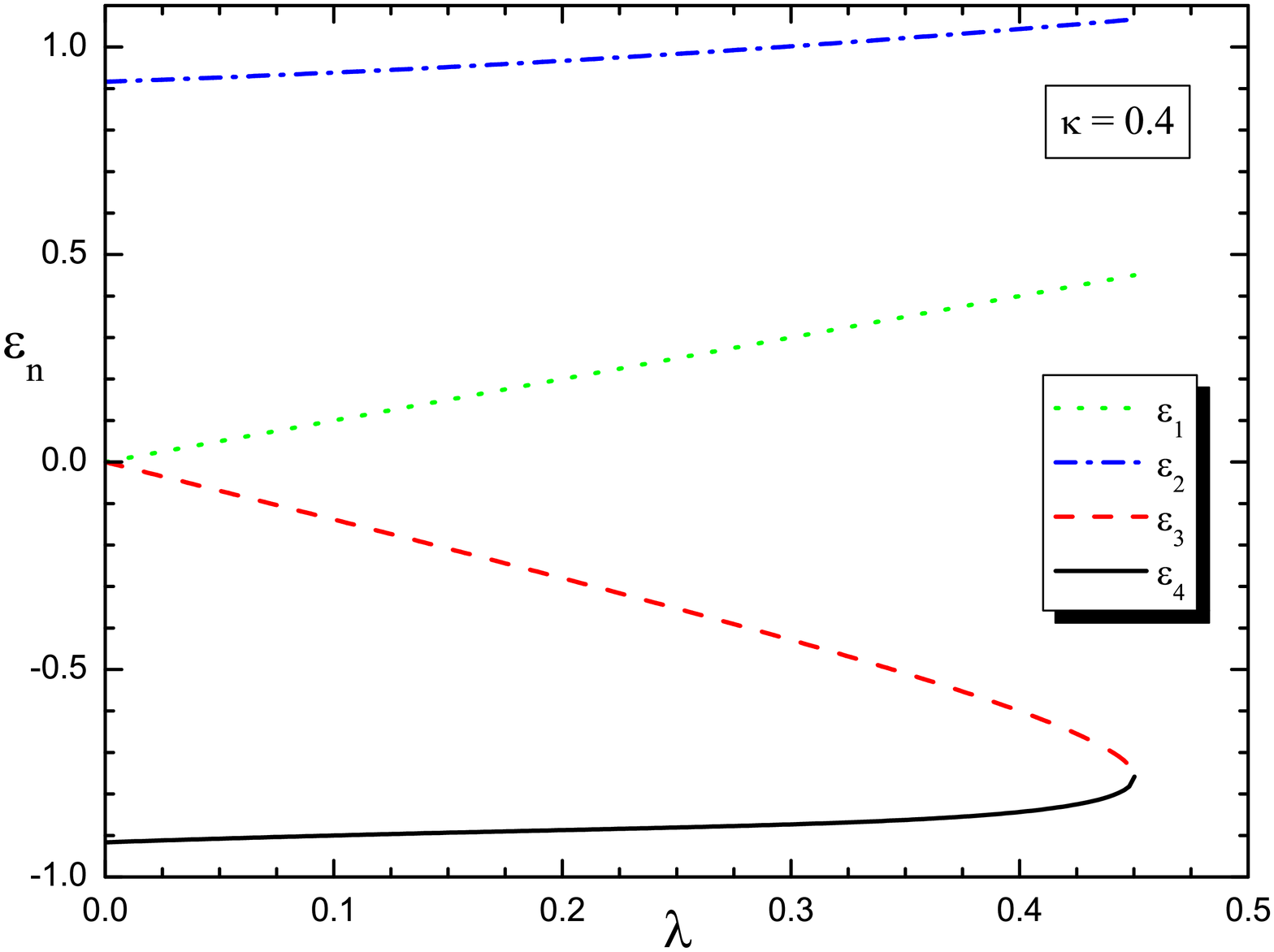}

\noindent {\small {\textbf{Figure 2:} Avoided level crossing: eigenvalues as
functions of $\lambda $ ($\kappa $) for fixed $\kappa $ ($\lambda $).}}

\medskip

\noindent where we observe the typical avoided level crossing behaviour of
the eigenvalues as a function of the parameters \cite{vNW}, i.e. the
eigenvalues $\varepsilon _{3}$ and $\varepsilon _{4}$ only meet in the
exceptional point when they simultaneously become complex.

For the computations of physical observables, which we will carry out below,
it is important to identify the lowest eigenvalue, which turns out to be
always $\varepsilon _{4}$.

Next we compute the right eigenvectors of $H(\lambda ,\kappa )$ to 
\begin{equation}
\left\vert \Phi _{1}\right\rangle =(0,-1,-1,0),\quad \left\vert \Phi
_{n}\right\rangle =(\gamma _{n},-\alpha _{n},-\alpha _{n},\beta _{n}),\quad
n=2,3,4,  \label{RE}
\end{equation}%
with $\alpha _{n}=i\kappa \left( \lambda -\varepsilon _{n}+1\right) $, $%
\beta _{n}=\kappa ^{2}+2\lambda ^{2}+2\lambda \varepsilon _{n}$ and $\gamma
_{n}=-\kappa ^{2}-2\varepsilon _{n}^{2}+2\lambda -2\lambda \varepsilon
_{n}+2\varepsilon _{n}$. We verify that left and right eigenvectors are
related via a conjugation $\left\vert \Psi _{n}\right\rangle =\left\langle
\Phi _{n}\right\vert $ and compute the signature as defined in (\ref{S}) to $%
s=(+,-,+,-)$ for the parity operator (\ref{PPP}). Normalizing the vectors in
(\ref{RE}) by dividing with $N_{1}=\sqrt{2}$, $N_{n}=(2\alpha _{n}^{2}+\beta
_{n}^{2}+\gamma _{n}^{2})^{1/2}$ for $n=2,3,4$ we compute the $\mathcal{C}$%
-operator according to (\ref{C}) to 
\begin{equation}
\mathcal{C}=\left( 
\begin{array}{cccc}
C_{5} & -C_{3} & -C_{3} & C_{4} \\ 
-C_{3} & -C_{1}-1 & -C_{1} & C_{2} \\ 
-C_{3} & -C_{1} & -C_{1}-1 & C_{2} \\ 
C_{4} & C_{2} & C_{2} & 2(C_{1}+1)-C_{5}%
\end{array}%
\right)  \label{CM}
\end{equation}%
where the matrix entries are 
\begin{equation}
\begin{array}{lll}
C_{1}=\frac{\alpha _{4}^{2}}{N_{4}^{2}}-\frac{\alpha _{2}^{2}}{N_{2}^{2}}-%
\frac{\alpha _{3}^{2}}{N_{3}^{2}}-\frac{1}{2},\quad & C_{2}=\frac{\alpha
_{4}\beta _{4}}{N_{4}^{2}}-\frac{\alpha _{2}\beta _{2}}{N_{2}^{2}}-\frac{%
\alpha _{3}\beta _{3}}{N_{3}^{2}},\quad & C_{3}=\frac{\alpha _{2}\gamma _{2}%
}{N_{2}^{2}}+\frac{\alpha _{3}\gamma _{3}}{N_{3}^{2}}-\frac{\alpha
_{4}\gamma _{4}}{N_{4}^{2}}, \\ 
C_{4}=\frac{\beta _{2}\gamma _{2}}{N_{2}^{2}}+\frac{\beta _{3}\gamma _{3}}{%
N_{3}^{2}}-\frac{\beta _{4}\gamma _{4}}{N_{4}^{2}}, & C_{5}=\frac{\gamma
_{2}^{2}}{N_{2}^{2}}+\frac{\gamma _{3}^{2}}{N_{3}^{2}}-\frac{\gamma _{4}^{2}%
}{N_{4}^{2}}. & 
\end{array}%
\end{equation}%
We may now verify that $\mathcal{C}$ indeed satisfied the properties (\ref%
{CP}) upon the use of the identities 
\begin{equation}
\begin{array}{lll}
C_{2}=C_{2}C_{5}-C_{3}C_{4}, & C_{3}=C_{5}C_{3}-C_{2}C_{4}-2C_{1}C_{3},\quad
& C_{4}=C_{2}C_{3}-C_{1}C_{4}, \\ 
1=2C_{3}^{2}+C_{4}^{2}+C_{5}^{2},\quad & 
0=C_{2}^{2}+C_{3}^{2}+2C_{1}(C_{1}+1). & 
\end{array}%
\end{equation}%
\quad Next we compute the metric operator in the form $\rho =\mathcal{PC}$
simply from (\ref{PPP}) and (\ref{CM}) to 
\begin{equation}
\rho =\left( 
\begin{array}{cccc}
C_{5} & -C_{3} & -C_{3} & C_{4} \\ 
C_{3} & 1+C_{1} & C_{1} & -C_{2} \\ 
C_{3} & C_{1} & 1+C_{1} & -C_{2} \\ 
C_{4} & C_{2} & C_{2} & 2(1+C_{1})-C_{5}%
\end{array}%
\right)
\end{equation}%
Since $i\alpha _{i},\beta _{i},\gamma _{i}\in \mathbb{R}$ it follows that $%
C_{1},iC_{2},iC_{3},C_{4},C_{5}\in \mathbb{R}$ and therefore we conclude
immediately that $\rho $ is Hermitian. To see whether $\rho $ is also
positive, as it ought to be, we compute its eigenvalues 
\begin{equation}
y_{1}=y_{2}=1\qquad \text{and\qquad }y_{3/4}=1+2C_{1}\pm 2\sqrt{%
C_{1}(1+C_{1})}.
\end{equation}%
Since $C_{1}>0$ all eigenvalues of $\rho $ are obviously guaranteed to be
positive.

Next we determine the corresponding eigenstates to 
\begin{equation}
\left\vert r_{1}\right\rangle =(0,-1,1,0),\quad \left\vert
r_{2}\right\rangle =(C_{4},0,0,1-C_{5}),\quad \left\vert
r_{3/4}\right\rangle =(\tilde{\gamma}_{3/4},\tilde{\alpha}_{3/4},\tilde{%
\alpha}_{3/4},\tilde{\beta}_{3/4})
\end{equation}
with $\tilde{\alpha}%
_{3/4}=y_{3/4}(C_{3}C_{4}+C_{2}(-4C_{1}+C_{5}-1))/2-C_{3}C_{4}$, $\tilde{%
\beta}_{3/4}=-C_{3}^{2}-C_{1}-C_{1}C_{5}+\left(
C_{3}^{2}+C_{1}(4C_{1}-C_{5}+3)\right) y_{3/4}$ and $\tilde{\gamma}%
_{3/4}=C_{1}C_{4}-C_{2}C_{3}+(C_{2}C_{3}+C_{1}C_{4})y_{3/4}$. Defining now
the matrix $U=\{r_{1},r_{2},r_{3},r_{4}\}$, whose column vectors are the
eigenvectors of $\rho $, we may take the square root of $\rho $, such that $%
\eta =\rho ^{1/2}=UD^{1/2}U^{-1}$, where $D=\limfunc{diag}%
(y_{1,}y_{2,},y_{3,},y_{4})$. The isospectral Hermitian counterpart of $H$
results from (\ref{1}) to an $XYZ$ spin chain (with just two sites) in a
magnetic field 
\begin{eqnarray}
h(\lambda ,\kappa ) &=&\eta H\eta ^{-1}=UD^{1/2}U^{-1}HUD^{-1/2}U^{-1}
\label{struc} \\
&=&\mu_{xx}^2(\lambda ,\kappa )S_{xx}^2+\mu _{yy}^2(\lambda ,\kappa
)S_{yy}^2+\mu _{zz}^2(\lambda ,\kappa )S_{zz}^2+\mu _{z}^2(\lambda ,\kappa
)S_z^2.  \label{mu}
\end{eqnarray}
It is clear that the coefficients $\mu _{xx}^2$, $\mu _{yy}^2$, $\mu _{zz}^2$%
, $\mu _{z}^2$ can be computed explicitly, but the expressions are rather
lengthy and we will therefore not present them here. They are all real
functions of $\lambda$ and $\kappa$. Their explicit form can be found in
appendix \ref{A} in terms of functions of $\lambda$ and $\kappa$ (\ref{ab})
which will be introduced in section \ref{5}, in the context of perturbation
theory. In the next section we wish to compare this exact result with a
perturbative expansion. Let us therefore report two numerical examples for
some isospectral Hermitian counterpart of $H(\lambda ,\kappa )$ 
\begin{equation}
h(0.1,0.5)=\left( 
\begin{array}{cccc}
-0.829536 & 0 & 0 & -0.0606492 \\ 
0 & -0.0341687 & -0.1341687 & 0 \\ 
0 & -0.1341687 & -0.0341687 & 0 \\ 
-0.0606492 & 0 & 0 & 0.897873%
\end{array}
\right) ,  \label{h1}
\end{equation}
and 
\begin{equation}
h(0.9,0.1)=\left( 
\begin{array}{cccc}
-0.985439 & 0 & 0 & -0.890532 \\ 
0 & -0.0094167 & -0.909417 & 0 \\ 
0 & -0.909417 & -0.0094167 & 0 \\ 
-0.890532 & 0 & 0 & 1.00427%
\end{array}
\right) .  \label{h2}
\end{equation}
Notice that $h_{23}=h_{32}=h_{22}-\lambda =h_{33}-\lambda $.

We have carried out a similar analysis for the chain with three sites
explicitly, albeit the resulting formulae are rather cumbersome to present.
In any case for longer chains one has to resort to more sophisticated and
less transparent techniques as for instance the Bethe ansatz. Alternatively,
we may employ perturbation theory.

\section{ The Yang-Lee quantum chain: perturbative results}

\label{5}

In this section we want to address the problem of obtaining the matrices $%
\eta $, $\rho $ and $h$ from a perturbative analysis as described in section %
\ref{pert}. We will study the $N=2,3$ and $4$ cases in detail and draw some
conclusions concerning the analytic expressions of $\eta $, $\rho $ and $h$
for generic $N$.

\subsection{ The $N=2$ case: perturbation theory in $\protect\kappa$}

\noindent Despite the fact that $H(\lambda ,\kappa )$ is just the $4\times 4$%
-matrix (\ref{H22}), it is actually not easy to find the matrix $q$ in (\ref%
{per}) exactly. As discussed in section \ref{ambi}, it is clear that the
equations (\ref{c1}) to (\ref{c5}) as well as the equations that would be
obtained for higher orders in perturbation theory, admit many solutions. Any
solution $q_{2k-1}$ can be modified by adding a matrix that commutes with $%
h_{0}(\lambda )$. However, not all solutions obtained in this manner would
be valid solutions if the equations (\ref{ecu}) are to hold. For the
particular case $N=2$, we are about to show that these constraints actually
select out a unique Hermitian counterpart to the Hamiltonian $H(\lambda
,\kappa )$. We will start by finding the most general matrix $q_{1}(\lambda
) $ which solves the identity (\ref{c1}). It is quite clear that given one
solution $q_{1}(\lambda )$, any matrix of the form $q_{1}(\lambda
)+B(\lambda )$ with $[h_{0}(\lambda ),B(\lambda )]=0$ will also be a
solution, so we may start by finding all such matrices. In this simple case,
there are four basic independent solutions to the equation $[h_{0}(\lambda
),B(\lambda )]=0$%
\begin{equation}
B_{1}=\mathbb{I}\text{,\quad }B_{2}=S_{zz}^{2}\text{,\quad }%
B_{3}=S_{xx}^{2}+S_{yy}^{2}\quad \text{and\quad }B_{4}=S_{z}^{2}-\lambda
S_{yy}^{2}\text{.}  \label{BB}
\end{equation}%
Since $h_{0}(\lambda )$ is a $4\times 4$-diagonalizable matrix, with
non-degenerate eigenvalues, there can be at most four independent matrices
that commute with it, namely those shown above or combinations thereof. On
the other hand, it is clear that any polynomial function of the Hamiltonian $%
h_{0}(\lambda )$ would also commute with $h_{0}(\lambda )$. As the four
matrices in (\ref{BB}) constitute a basis, we expect to be able to express
any power of $h_{0}$ as linear combinations of them. Indeed, we find 
\begin{eqnarray}
h_{0}(\lambda )^{2n} &=&\frac{(1+\lambda ^{2})^{n}}{2}(B_{1}+\frac{1}{2}%
B_{2})+\frac{\lambda ^{2n}}{2}(B_{1}-\frac{1}{2}B_{2}), \\
h_{0}(\lambda )^{2n+1} &=&(1+\lambda ^{2})^{n}h_{0}(\lambda )+\frac{\lambda
(1+\lambda ^{2})^{n}-\lambda ^{2n+1}}{4}B_{3},
\end{eqnarray}%
for $n\in \mathbb{N}_{0}$. Therefore, the most general solution to the first
order equation (\ref{c1}) for the present model is 
\begin{equation}
q_{1}(\lambda )=-S_{y}^{2}-\lambda
(S_{yz}^{2}+S_{zy}^{2})+\sum\limits_{i=1}^{4}{{f_{i}(\lambda )}}B_{i},
\label{A11}
\end{equation}%
where the $f_{i}(\lambda )$, $i=1,2,3,4$ are arbitrary functions of $\lambda 
$.

Before we proceed to determine $q_{3}(\lambda )$ by solving (\ref{c3}) let
us comment on the ambiguities and answer the question of whether all
solutions (\ref{A11}) are compatible with the equations (\ref{ecu}).
Specializing equations (\ref{pmat}) and (\ref{tmat}) for the matrices in (%
\ref{A11}) we find 
\begin{equation}
\mathcal{P}X\mathcal{P=\mathcal{T}}X\mathcal{\mathcal{T=}-}X,\qquad \text{%
for }X=S_{y}^{2},S_{yz}^{2},S_{zy}^{2}
\end{equation}%
whereas 
\begin{equation}
\mathcal{P}B_{i}\mathcal{P}=\mathcal{T}B_{i}\mathcal{T}=B_{i},\qquad \text{%
for }i=1,2,3,4.  \label{ide}
\end{equation}%
These equations imply that the equalities (\ref{ecu}) can only be satisfied
if the functions $f_{i}(\lambda )=0$ for $i=1,2,3,4$. Thus we have selected
out a \textit{unique} solution for $q_{1}(\lambda )$, namely 
\begin{equation}
q_{1}(\lambda )=-S_{y}^{2}-\lambda (S_{yz}^{2}+S_{zy}^{2}).  \label{uni1}
\end{equation}%
More generally, the conditions (\ref{ecu}) together with the properties (\ref%
{pmat}) and (\ref{tmat}) imply that

\begin{itemize}
\item any solutions $q_{2k-1}$ must be linear combinations of matrices (\ref%
{matrices}) with $n_y$ odd,

\item any solutions $q_{2k-1}$ must be linear combinations of matrices (\ref%
{matrices}) with $n_y+n_x$ odd,

\item or, combining the two conditions above, any solutions $q_{2k-1}$ must
be linear combinations of matrices (\ref{matrices}) with $n_y$ odd and $n_x$
even,
\end{itemize}

as anticipated in subsection \ref{unih00}. These conditions then
automatically guarantee the validity of the $\mathcal{PT}$-properties (\ref%
{cons}) for the $q_{2k-1}$. For $N=2$, this singles out the matrices $%
S_{y}^{2}$ and $S_{yz}^{2}=S_{zy}^{2}$ in (\ref{uni1}), so that, even before
attempting to solve (\ref{c1}) we would already know that it can only be a
linear combination of those two matrices. As indicated above, these
constraints apply for all other $q_{2k-1}(\lambda )$, with $k>1$ so that we
can safely claim that, at all orders in perturbation theory, the matrices $%
q_{2k-1}(\lambda )$ must be linear combinations of the form, 
\begin{equation}
q_{2k-1}(\lambda )=a_{2k-1}(\lambda )S_{y}^{2}+b_{2k-1}(\lambda
)(S_{yz}^{2}+S_{zy}^{2}),  \label{ak}
\end{equation}
where $a_{2k-1}(\lambda ),b_{2k-1}(\lambda )$ are real functions of $\lambda 
$. In other words, all the terms in the perturbative expansion of $q$ are
linear combinations of the same two matrices. Hence, we can write 
\begin{equation}
e^{q}=e^{\alpha (\lambda ,\kappa )S_{y}^{2}+\beta (\lambda ,\kappa
)(S_{yz}^{2}+S_{zy}^{2})},  \label{oi}
\end{equation}
which, after computing the exponential becomes 
\begin{equation*}
\left( 
\begin{array}{cccc}
\frac{\rho (\lambda ,\kappa )^{2}+\epsilon (\lambda ,\kappa )^{2}\cosh
[2\gamma (\lambda ,\kappa )]}{2\gamma (\lambda ,\kappa )^{2}} & -\frac{%
i\epsilon (\lambda ,\kappa )\sinh [2\gamma (\lambda ,\kappa )]}{2\gamma
(\lambda ,\kappa )} & -\frac{i\epsilon (\lambda ,\kappa )\sinh [2\gamma
(\lambda ,\kappa )]}{2\gamma (\lambda ,\kappa )} & -\frac{\delta (\lambda
,\kappa )\sinh ^{2}[\gamma (\lambda ,\kappa )]}{\gamma (\lambda ,\kappa )^{2}%
} \\ 
\frac{i\epsilon (\lambda ,\kappa )\sinh [2\gamma (\lambda ,\kappa )]}{%
2\gamma (\lambda ,\kappa )} & \cosh ^{2}\gamma (\lambda ,\kappa ) & \sinh
^{2}\gamma (\lambda ,\kappa ) & -\frac{i\rho (\lambda ,\kappa )\sinh
[2\gamma (\lambda ,\kappa )]}{2\gamma (\lambda ,\kappa )} \\ 
\frac{i\epsilon (\lambda ,\kappa )\sinh [2\gamma (\lambda ,\kappa )]}{%
2\gamma (\lambda ,\kappa )} & \sinh ^{2}\gamma (\lambda ,\kappa ) & \cosh
^{2}\gamma (\lambda ,\kappa ) & -\frac{i\rho (\lambda ,\kappa )\sinh
[2\gamma (\lambda ,\kappa )]}{2\gamma (\lambda ,\kappa )} \\ 
\frac{\delta (\lambda ,\kappa )\sinh ^{2}\gamma (\lambda ,\kappa )}{\gamma
(\lambda ,\kappa )^{2}} & \frac{i\rho (\lambda ,\kappa )\sinh [2\gamma
(\lambda ,\kappa )]}{2\gamma (\lambda ,\kappa )} & \frac{i\rho (\lambda
,\kappa )\sinh [2\gamma (\lambda ,\kappa )]}{2\gamma (\lambda ,\kappa )} & 
\frac{\epsilon (\lambda ,\kappa )^{2}+\rho (\lambda ,\kappa )^{2}\cosh
[2\gamma (\lambda ,\kappa )]}{2\gamma (\lambda ,\kappa )^{2}}%
\end{array}
\right)
\end{equation*}
where 
\begin{equation}
\alpha (\lambda ,\kappa )=\sum_{k=0}^{\infty }\kappa ^{2k+1}a_{2k+1}(\lambda
),\qquad \beta (\lambda ,\kappa )=\sum_{k=0}^{\infty }\kappa
^{2k+1}b_{2k+1}(\lambda ),  \label{ab}
\end{equation}
and 
\begin{equation}
\gamma (\lambda ,\kappa )=\sqrt{\alpha (\lambda ,\kappa )^{2}+4\beta
(\lambda ,\kappa )^{2}},\quad \delta (\lambda ,\kappa )=\alpha (\lambda
,\kappa )^{2}-4\beta (\lambda ,\kappa )^{2}.  \label{ga1}
\end{equation}
\begin{equation}
\epsilon (\lambda ,\kappa )=\alpha (\lambda ,\kappa )+2\beta (\lambda
,\kappa ),\quad \rho (\lambda ,\kappa )=\alpha (\lambda ,\kappa )-2\beta
(\lambda ,\kappa ).  \label{ga2}
\end{equation}
Notice that, for $\alpha (\lambda ,\kappa )$ and $\beta (\lambda ,\kappa )$
real, the matrix above is explicitly Hermitian, as it should be. Once the
coefficients $\alpha (\lambda ,\kappa )$ and $\beta (\lambda ,\kappa )$ have
been obtained, the Hermitian Hamiltonian (\ref{1}) can be easily computed.
The difficulty here is however that general formulae for the coefficients $%
a_{2k+1}(\lambda )$ and $b_{2k+1}(\lambda )$ are very difficult to obtain.
Nonetheless, perturbation theory allows us to compute these coefficients up
to very high orders in powers of $\kappa $. In order to solve for such high
orders, we have resorted to the use of the algebraic manipulation software
Mathematica. It allows us to find the entries of the matrix (\ref{oi}) as
perturbative series in $\kappa $ and to fix the coefficients $%
a_{2k+1}(\lambda )$ and $b_{2k+1}(\lambda )$ by matching the entries of $%
H^{\dagger }(\lambda ,\kappa )$ and $\eta ^{2}H(\lambda ,\kappa )\eta ^{-2}$%
, order by order in perturbation theory, as expected from (\ref{1}). For
numerical computations and sufficiently small values of $\kappa $ this gives
results which are very close to the exact values. In tables 1 and 2 we
present the coefficients $a_{2k+1}(\lambda )$ and $b_{2k+1}(\lambda )$ up to 
$k=7$.

\medskip

{\small \noindent 
\begin{tabular}{|c||c|c|c|c|c|c|c|c|}
\hline
& $-\lambda^0$ & $-\lambda^2$ & $-\lambda^4$ & $-\lambda^6$ & $-\lambda^8$ & 
$-\lambda^{10}$ & $-\lambda^{12}$ & $-\lambda^{14}$ \\ \hline\hline
$a_1 (\lambda)$ & 1 & 0 & 0 & 0 & 0 & 0 & 0 & 0 \\ \hline
$a_3 (\lambda)$ & $\frac{1}{3}$ & $\frac{2^4}{3}$ & 0 & 0 & 0 & 0 & 0 & 0 \\ 
\hline
$a_5 (\lambda)$ & $\frac{1}{5}$ & $\frac{244}{15}$ & $\frac{2^8}{5}$ & 0 & 0
& 0 & 0 & 0 \\ \hline
$a_7 (\lambda)$ & $\frac{1}{7}$ & $\frac{1152}{35}$ & $\frac{35104}{105}$ & $%
\frac{2^{12}}{7}$ & 0 & 0 & 0 & 0 \\ \hline
$a_9 (\lambda)$ & $\frac{1}{9}$ & $\frac{17432}{315}$ & $\frac{43408}{35}$ & 
$\frac{1890368}{315}$ & $\frac{2^{16}}{9}$ & 0 & 0 & 0 \\ \hline
$a_{11} (\lambda)$ & $\frac{1}{11}$ & $\frac{289616}{3465}$ & $\frac{797296}{%
231}$ & $\frac{38228224}{1155}$ & $\frac{355526144}{3465}$ & $\frac{2^{20}}{%
11}$ & 0 & 0 \\ \hline
$a_{13} (\lambda)$ & $\frac{1}{13}$ & $\frac{353372}{3003}$ & $\frac{72293440%
}{9009}$ & $\frac{655729408}{5005}$ & $\frac{2275245568}{3003}$ & $\frac{%
15442769920}{9009}$ & $\frac{2^{24}}{13}$ & 0 \\ \hline
$a_{15} (\lambda)$ & $\frac{1}{15}$ & $\frac{7100416}{45045}$ & $\frac{%
67453952}{4095}$ & $\frac{896579072}{2145}$ & $\frac{58903814656}{15015}$ & $%
\frac{717363822592}{45045} $ & $\frac{1273503367168}{45045}$ & $\frac{2^{28}%
}{15}$ \\ \hline
\end{tabular}
}

\noindent {\small {\textbf{Table 1:} The coefficients ${a}_{2k+1}(\lambda )$
for $k <8$. }}

\medskip

{\small \noindent 
\begin{tabular}{|c||c|c|c|c|c|c|c|c|}
\hline
& $-\lambda$ & $-\lambda^3$ & $-\lambda^5$ & $-\lambda^7$ & $-\lambda^9$ & $%
-\lambda^{11}$ & $-\lambda^{13}$ & $-\lambda^{15}$ \\ \hline\hline
$b_1(\lambda)$ & 1 & 0 & 0 & 0 & 0 & 0 & 0 & 0 \\ \hline
$b_3(\lambda)$ & $\frac{4}{3}$ & $\frac{2^4}{3}$ & 0 & 0 & 0 & 0 & 0 & 0 \\ 
\hline
$b_5(\lambda)$ & $\frac{23}{15} $ & $\frac{2^7}{5}$ & $\frac{2^8}{5}$ & 0 & 0
& 0 & 0 & 0 \\ \hline
$b_7(\lambda)$ & $\frac{176}{105}$ & $\frac{7544}{105}$ & $\frac{2^{10}(3)}{7%
} $ & $\frac{2^{12}}{7}$ & 0 & 0 & 0 & 0 \\ \hline
$b_9(\lambda)$ & $\frac{563}{315}$ & $\frac{49136}{315}$ & $\frac{212816}{105%
}$ & $\frac{2^{16}}{9}$ & $\frac{2^{16}}{9}$ & 0 & 0 & 0 \\ \hline
$b_{11}(\lambda)$ & $\frac{6508}{3465}$ & $\frac{335576}{1155}$ & $\frac{%
7827328}{1155} $ & $\frac{164005504}{3465}$ & $\frac{2^{18} (5) }{11}$ & $%
\frac{2^{20}}{11}$ & 0 & 0 \\ \hline
$b_{13}(\lambda) $ & $\frac{88069}{45045}$ & $\frac{4400960}{9009}$ & $\frac{%
39578944}{2145}$ & $\frac{1947324416}{9009}$ & $\frac{9037578752}{9009}$ & $%
\frac{2^{23}(3)}{13}$ & $\frac{2^{24}}{13}$ & 0 \\ \hline
$b_{15}(\lambda)$ & $\frac{91072}{45045}$ & $\frac{34381136}{45045}$ & $%
\frac{178162048}{4095}$ & $\frac{1068366848}{1365}$ & $\frac{37570428928}{%
6435}$ & $\frac{903387164672}{45045}$ & $\frac{2^{26}(7)}{15}$ & $\frac{%
2^{28}}{15}$ \\ \hline
\end{tabular}
}

\medskip

\noindent {\small {\textbf{Table 2:} The coefficients ${b}_{2k+1}(\lambda )$
for $k<8$.}}

\medskip

\noindent These tables should be understood as follows: in order to obtain
the corresponding coefficient the numbers in a given row are to be
multiplied by the power of $\lambda $ (with a minus sign added) at the top
of the same column and added up. For example: 
\begin{equation}
a_{5}(\lambda )=-\frac{1}{5}-\frac{244{\lambda }^{2}}{15}-\frac{2^{8}{%
\lambda }^{4}}{5}.
\end{equation}

The only case for which it is easy to conjecture the expressions of $%
a_{2k+1}(\lambda ),b_{2k+1}(\lambda )$ for generic values of $k$ corresponds
to $\lambda =0$. Then $a_{2k+1}(0)=-1/(2k+1)$ and $b_{2k+1}(0)=0$, which
gives the already known result $\alpha (0,\kappa )=-\text{arctanh}(\kappa )$
and $\beta (0,\kappa )=0$, see section \ref{known}. Having found $\eta $, it
is straightforward using (\ref{1}) to determine the Hermitian counterpart of 
$H(\lambda ,\kappa )$. In general, we find 
\begin{eqnarray}
&&h(\lambda ,\kappa )=e^{q/2}H(\lambda ,\kappa )e^{-q/2}=\left( 
\begin{array}{cccc}
h_{11}(\lambda ,\kappa ) & 0 & 0 & h_{14}(\lambda ,\kappa ) \\ 
0 & h_{22}(\lambda ,\kappa ) & h_{22}(\lambda ,\kappa )-\lambda & 0 \\ 
0 & h_{22}(\lambda ,\kappa )-\lambda & h_{22}(\lambda ,\kappa ) & 0 \\ 
h_{14}(\lambda ,\kappa ) & 0 & 0 & h_{44}(\lambda ,\kappa )%
\end{array}%
\right)  \notag \\
&=&\frac{h_{22}(\lambda ,\kappa )-\lambda +h_{14}(\lambda ,\kappa )}{4}%
S_{xx}^{2}+\frac{h_{22}(\lambda ,\kappa )-\lambda -h_{14}(\lambda ,\kappa )}{%
4}S_{yy}^{2}  \notag \\
&&+\frac{h_{11}(\lambda ,\kappa )+h_{44}(\lambda ,\kappa )-2h_{22}(\lambda
,\kappa )}{8}S_{zz}^{2}+\frac{h_{11}(\lambda ,\kappa )-h_{44}(\lambda
,\kappa )}{4}S_{z}^{2}  \notag \\
&&+\frac{h_{11}(\lambda ,\kappa )+h_{44}(\lambda ,\kappa )+2h_{22}(\lambda
,\kappa )}{4},  \label{strucc}
\end{eqnarray}%
which is the same kind of structure found in (\ref{mu}). The functions $%
h_{11}(\lambda ,\kappa )$, $h_{22}(\lambda ,\kappa )$, $h_{14}(\lambda
,\kappa )$ and $h_{44}(\lambda ,\kappa )$ are real functions of the coupling
constants which can be evaluated very accurately for fixed values of $%
\lambda $ and $\kappa $ by using the perturbative results above. In fact,
the remaining entries of the matrix are not explicitly zero as functions of $%
\alpha (\lambda ,\kappa )$ and $\beta (\lambda ,\kappa )$. They are
complicated functions of the latter which when carrying out the perturbation
theory result to be zero up order $\kappa ^{15}$. This is consistent with
the exact results obtained before. The explicit expressions of the entries
of $h(\lambda ,\kappa )$ in terms of the functions (\ref{ga1}) and (\ref{ga2}%
) can be found in appendix \ref{A}. Here, we will just present their
expression as a series expansion in $\kappa $ up to order $\kappa ^{4}$ (for
higher orders, expression become too cumbersome), 
\begin{eqnarray}
h_{11}(\lambda ,\kappa ) &=&-1+\left( \frac{1}{2}+\lambda \right) {\kappa }%
^{2}+\left( \frac{1}{8}+\lambda +\frac{3{\lambda }^{2}}{2}+4{\lambda }%
^{3}\right) {\kappa }^{4}+\mathcal{O}(\kappa ^{6}), \\
h_{22}(\lambda ,\kappa ) &=&-\lambda {\kappa }^{2}-\lambda \left( 1+4{%
\lambda }^{2}\right) {\kappa }^{4}+\mathcal{O}(\kappa ^{6}), \\
h_{44}(\lambda ,\kappa ) &=&1+\left( -\frac{1}{2}+\lambda \right) {\kappa }%
^{2}+\left( -\frac{1}{8}+\lambda -\frac{3{\lambda }^{2}}{2}+4{\lambda }%
^{3}\right) {\kappa }^{4}+\mathcal{O}(\kappa ^{6}), \\
h_{14}(\lambda ,\kappa ) &=&-\lambda +\lambda {\kappa }^{2}+\left( \frac{%
3\lambda }{2}+4{\lambda }^{3}\right) {\kappa }^{4}+\mathcal{O}(\kappa ^{6}).
\end{eqnarray}%
From this expansions we can deduce some interesting features which also
extend to higher orders in perturbation theory 
\begin{equation}
h_{11}(-\lambda ,\kappa )=-h_{44}(\lambda ,\kappa ),\quad h_{22}(-\lambda
,\kappa )=-h_{22}(\lambda ,\kappa ),\quad h_{14}(-\lambda ,\kappa
)=-h_{14}(\lambda ,\kappa ).  \label{1144}
\end{equation}%
Furthermore, we note that the Hermitian Hamiltonian $h(\lambda ,\kappa )$ is
an even function of $\kappa $, so that the series expansion of its
components involves only even powers of the coupling. Finally, as it should
be, the Hamiltonian $h(\lambda ,\kappa )$ is also $\mathcal{PT}$-symmetric,
which follows from the fact that all matrices involved ($%
S_{xx}^{2},S_{yy}^{2},S_{zz}^{2}$ and $S_{z}^{2}$) are invariant under the
adjoint action of the operator $\mathcal{PT}$. These are in fact the only
matrices that are both $\mathcal{PT}$-symmetric and real. In fact we could
have known a priori before carrying any computations that $h(\lambda ,\kappa
)$ has to be some linear combination of $S_{xx}^{2},S_{yy}^{2},S_{zz}^{2}$
and $S_{z}^{2}$. Notice that the reality of $h(\lambda ,\kappa )$ can be
expressed by saying that any matrices (\ref{matrices}) involved must have $%
n_{y}$ even, as defined in the paragraph after equation (\ref{ptmat}).

In order to compare with the results obtained in section \ref{4} we give
below the numerical values of the entries of the Hermitian Hamiltonian $%
h(\lambda ,\kappa )$ for fixed values of the couplings 
\begin{equation}
h(0.1,0.5)=\left( 
\begin{array}{cccc}
-0.82953\underline{4} & 0 & 0 & -0.0606\underline{716} \\ 
0 & -0.034168\underline{8} & -0.13416\underline{9} & 0 \\ 
0 & -0.13416\underline{9} & -0.034168\underline{8} & 0 \\ 
-0.0606\underline{716} & 0 & 0 & 0.89787\underline{2}%
\end{array}%
\right) ,  \label{15}
\end{equation}%
and 
\begin{equation}
h(0.9,0.1)=\left( 
\begin{array}{cccc}
-0.985439 & 0 & 0 & -0.890532 \\ 
0 & -0.00941674 & -0.909417 & 0 \\ 
0 & -0.909417 & -0.00941674 & 0 \\ 
-0.890532 & 0 & 0 & 1.00427%
\end{array}%
\right) .  \label{91}
\end{equation}%
We underlined the digits which differ from the exact values computed in (\ref%
{h1}) and (\ref{h2}) and note that the perturbative expressions for $%
h(0.1,0.5)$ and $h(0.9,0.1)$ agree extremely well with them, especially for
smaller values of $\kappa $, as is expected.

In order to see how fast this precision is reached in the perturbation
theory we report in table 3 the relative error for the entry $h_{11}$ order
by order up to $15$

\medskip

\noindent 
\begin{tabular}{|l||c|c|c|c|c|c|c|}
\hline
$\lambda ,\kappa \backslash \mathcal{O}(\kappa )$ & $2$ & $4$ & $6$ & $8$ & $%
10$ & $12$ & $14$ \\ \hline\hline
$0.9,0.1$ & $5.7~10^{-4}$ & $4.6~10^{-5}$ & $4.7~10^{-6}$ & $5.3~10^{-7}$ & $%
6.4~10^{-8}$ & $8.2~10^{-9}$ & $1.1~10^{-9}$ \\ \hline
$0.1,0.5$ & $2.5~10^{-2}$ & $6.3~10^{-3}$ & $2.1~10^{-3}$ & $7.5~10^{-4}$ & $%
2.9~10^{-4}$ & $1.6~10^{-4}$ & $4.7~10^{-5}$ \\ \hline
\end{tabular}

\medskip

\noindent {\small {\textbf{Table 3:} Relative error = |(perturbative value -
exact value) / exact value| for $h_{11}$ order by order.}}

\medskip

\noindent We observe that the convergence is fairly fast, which allows to
extract useful information from the perturbation theory even at low order.
We shall not be concerned here with more rigorous mathematical arguments
regarding the summability and convergence in general.

\subsection{The $N=2$ case: perturbation theory in $\protect\lambda$}

In the previous section we have employed the standard version of
perturbation theory when dealing with non-Hermitian Hamiltonians of the type
(\ref{H}), that is decomposing the Hamiltonian into a Hermitian and a
non-Hermitian part as in (\ref{h0h1}) and then treating the non-Hermitian
part as the perturbation. Since the Hamiltonian (\ref{H}) depends on two
independent coupling constants, $\kappa $ and $\lambda $, it is also
natural, albeit less standard, to consider perturbation theory in the
coupling constant $\lambda $ rather than in $\kappa $. In other words we
expand around the exact solution for $\lambda =0$ provided in section $4.1$
and treat the nearest neighbour interaction term as perturbation. As
announced already in section 3.2., we decompose $H(\lambda ,\kappa )$ into 
\begin{equation}
H(\lambda ,\kappa )=\tilde{H}_{0}(\kappa )+\lambda \tilde{h}_{1},\quad \text{%
where}\quad \tilde{H}_{0}(\kappa )=-\frac{1}{2}(S_{z}^{N}+i\kappa
S_{x}^{N}),\qquad \tilde{h}_{1}=-\frac{1}{2}S_{xx}^{N}.
\end{equation}%
We wish now once again to solve the equations (\ref{1}) for the Dyson map $%
\eta $, that is 
\begin{equation}
H^{\dagger }(\lambda ,\kappa )=e^{w}H(\lambda ,\kappa )e^{-w},  \label{ag}
\end{equation}%
where we have assumed that $\eta $ admits the exponential form 
\begin{equation}
\eta =e^{w/2}\quad \qquad \text{with}\quad \qquad w=\sum_{a=0}^{\infty
}\lambda ^{a}w_{a}(\kappa ).  \label{exp}
\end{equation}%
At order $\lambda ^{0}$ equation (\ref{ag}) becomes simply 
\begin{equation}
\tilde{H}_{0}^{\dagger }(\kappa )=e^{w_{0}(\kappa )}\tilde{H}_{0}(\kappa
)e^{-w_{0}(\kappa )}.  \label{order0}
\end{equation}%
The solution to this equation for all $N$ was found in subsection \ref{limit}
and corresponds to the Dyson map identified in equation (\ref{w0}). For $N=2$
this means that 
\begin{equation}
w_{0}(\kappa )=-\text{arctanh}(\kappa )S_{y}^{2}.  \label{w00}
\end{equation}%
Employing the once again the Backer-Campbell-Hausdorff identity to select $%
\mathcal{O}(\lambda )$ terms in (\ref{ag}) we find the condition 
\begin{equation}
\tilde{h}_{1}=e^{w_{0}(\kappa )}\tilde{h}_{1}e^{-w_{0}(\kappa
)}+\sum_{k=1}^{\infty }\sum_{i=1}^{k}\sum_{\substack{ a_{i}=1,  \\ a_{j\neq
i}=0}}\frac{1}{k!}\left[ w_{a_{1}}(\kappa ),\left[ w_{a_{2}}(\kappa ),\cdots
,\left[ w_{a_{k}}(\kappa ),H_{0}(\kappa )\right] \cdots \right] \right]
\label{order11}
\end{equation}%
Notice that, because of the presence of the zeroth order term $w_{0}(\kappa
) $, the equation (\ref{order11}) involves a sum of infinitely many
contributions, as would equations corresponding to higher orders in
perturbation theory. Because of this, it would in general be difficult to
solve (\ref{ag}) using perturbation theory in $\lambda $. However, for $N=2$
we can solve up to high orders in $\lambda $ by exploiting the fact that $%
\eta $ must have the structure identified in the previous section. This
means that $\eta $ is a matrix of the form (\ref{oi}) with 
\begin{equation}
\alpha (\lambda ,\kappa )=\sum_{a=0}^{\infty }\lambda ^{a}y_{a}(\kappa
),\qquad \beta (\lambda ,\kappa )=\sum_{a=0}^{\infty }\lambda
^{a}z_{a}(\kappa ).  \label{albe}
\end{equation}%
It is then possible to find the real functions $y_{a}(\kappa )$ and $%
z_{a}(\kappa )$ which solve equation (\ref{ag}) order by order in $\lambda $
by employing Mathematica, as explained in the previous subsection. In this
way, we have obtained the functions $y_{a}(\kappa )$ and $z_{a}(\kappa )$
above up to order $\lambda ^{15}$. Here we will just report the first five
orders, 
\begin{eqnarray}
y_{0}(\kappa ) &=&-\text{arctanh}(\kappa ), \\
z_{1}(\kappa ) &=&\frac{y_{0}(\kappa )}{1-\kappa ^{2}}, \\
y_{2}(\kappa ) &=&-\frac{2(\kappa +2{\kappa }^{3}+\left( 1-{\kappa }%
^{2}\right) y_{0}(\kappa ))}{{\left( 1-{\kappa }^{2}\right) }^{3}}, \\
z_{3}(\kappa ) &=&-\frac{2(\kappa +2{\kappa }^{3}+\left( 1-{\kappa }^{2}-2{%
\kappa }^{4}\right) y_{0}(\kappa ))}{{\left( 1-{\kappa }^{2}\right) }^{4}},
\\
y_{4}(\kappa ) &=&\frac{2\left( \kappa \left( 3-5{\kappa }^{2}-32{\kappa }%
^{4}-8{\kappa }^{6}\right) +\left( 3-6{\kappa }^{2}-5{\kappa }^{4}+8{\kappa }%
^{6}\right) y_{0}(\kappa )\right) }{{\left( 1-{\kappa }^{2}\right) }^{6}}, \\
z_{5}(\kappa ) &=&\frac{2\left( \kappa \left( 3-5{\kappa }^{2}-36{\kappa }%
^{4}-16{\kappa }^{6}\right) +\left( 3-6{\kappa }^{2}-9{\kappa }^{4}+28{%
\kappa }^{6}+8{\kappa }^{8}\right) y_{0}(\kappa )\right) }{{\left( 1-{\kappa 
}^{2}\right) }^{7}},
\end{eqnarray}%
and $y_{2a+1}(\kappa )=z_{2a}(\kappa )=0$ for all $a=0,1,\ldots $ From these
formulae, it is possible to find an expression for the Hermitian Hamiltonian 
$h(\lambda ,\kappa )$ as a perturbative series in $\lambda $. As it should
be, one finds the same structure (\ref{strucc}) with 
\begin{eqnarray}
&&h_{11}(\lambda ,\kappa ) =-\sqrt{1-{\kappa }^{2}}+\frac{{\kappa }%
^{2}\lambda }{1-{\kappa }^{2}}-\frac{6\left( -2+{\kappa }^{2}+2\sqrt{1-{%
\kappa }^{2}}\right) {\lambda }^{2}}{{\left( 1-{\kappa }^{2}\right) }^{\frac{%
5}{2}}}+\frac{4{\kappa }^{4}{\lambda }^{3}}{{\left( 1-{\kappa }^{2}\right) }%
^{4}}  \notag \\
&&-\frac{2\left( 40-44{\kappa }^{2}-57{\kappa }^{4}+28{\kappa }^{6}+8\sqrt{1-%
{\kappa }^{2}}\left( -5+3{\kappa }^{2}+8{\kappa }^{4}\right) \right) {%
\lambda }^{4}}{{\left( 1-{\kappa }^{2}\right) }^{\frac{11}{2}}}+\mathcal{O}%
(\lambda ^{5}),  \label{hhh11} \\
&&h_{22}(\lambda ,\kappa ) =-\frac{{\kappa }^{2}\lambda }{1-{\kappa }^{2}}-%
\frac{4{\kappa }^{4}{\lambda }^{3}}{{\left( 1-{\kappa }^{2}\right) }^{4}}+%
\mathcal{O}(\lambda ^{5}), \\
&&h_{44}(\lambda ,\kappa ) =\sqrt{1-{\kappa }^{2}}+\frac{{\kappa }%
^{2}\lambda }{1-{\kappa }^{2}}+\frac{6\left( -2+{\kappa }^{2}+2\sqrt{1-{%
\kappa }^{2}}\right) {\lambda }^{2}}{{\left( 1-{\kappa }^{2}\right) }^{\frac{%
5}{2}}}+\frac{4{\kappa }^{4}{\lambda }^{3}}{{\left( 1-{\kappa }^{2}\right) }%
^{4}}  \notag \\
&&+\frac{2\left( 40-44{\kappa }^{2}-57{\kappa }^{4}+28{\kappa }^{6}+8\sqrt{1-%
{\kappa }^{2}}\left( -5+3{\kappa }^{2}+8{\kappa }^{4}\right) \right) {%
\lambda }^{4}}{{\left( 1-{\kappa }^{2}\right) }^{\frac{11}{2}}}+\mathcal{O}%
(\lambda ^{5}), \\
&&h_{14}(\lambda ,\kappa ) =\frac{\left( -4+4{\kappa }^{2}+3\sqrt{1-{\kappa }%
^{2}}\right) \lambda }{{\left( 1-{\kappa }^{2}\right) }^{\frac{3}{2}}} 
\notag \\
&&+\frac{4\left( 8-10{\kappa }^{2}-2{\kappa }^{4}+4{\kappa }^{6}+\sqrt{1-{%
\kappa }^{2}}\left( 2+{\kappa }^{2}\right) \left( -4+5{\kappa }^{2}\right)
\right) {\lambda }^{3}}{{\left( 1-{\kappa }^{2}\right) }^{\frac{9}{2}}}+%
\mathcal{O}(\lambda ^{5}).  \label{hhh14}
\end{eqnarray}%
Notice that the same symmetries (\ref{1144}) are also found here. We also
see once again that $h(\lambda ,\kappa )$ is an even function of $\kappa $,
as only even powers are involved. Computing again numerical values for $%
h(0.1,0.5)$ and $h(0.9,0.1)$ we find almost perfect agreement with the exact
results. There is extremely good agreement both with the exact results (\ref%
{h1}) and (\ref{h2}) and with the result from perturbation theory in $\kappa 
$ (\ref{15}) and (\ref{91}). In order to see how fast this precision is
reached in the perturbation theory we report in table 4 the relative error
for the entry $h_{11}$order by order up to $15$, omitting the odd orders
despite the fact that they occur in the $\lambda $-perturbation theory

\medskip

\noindent 
\begin{tabular}{||l||c|c|c|c|c|c|c||}
\hline
$\lambda ,\kappa \backslash \mathcal{O(\lambda )}$ & $2$ & $4$ & $6$ & $8$ & 
$10$ & $12$ & $14$ \\ \hline
$0.9,0.1$ & $3.4~10^{-3}$ & $2.3~10^{-5}$ & $1.9~10^{-6}$ & $1.8~10^{-7}$ & $%
1.9~10^{-8}$ & $2.0~10^{-9}$ & $2.2~10^{-10}$ \\ \hline
$0.1,0.5$ & $1.1~10^{-3}$ & $6.3~10^{-5}$ & $4.9~10^{-6}$ & $3.6~10^{-7}$ & $%
3.1~10^{-8}$ & $2.8~10^{-9}$ & $2.6~10^{-10}$ \\ \hline
\end{tabular}

\medskip

\noindent {\small {\textbf{Table 4:} Relative error = |(perturbative value -
exact value) / exact value| for $h_{11}$ order by order.}}

\medskip

\noindent We note that the perturbation theory converges extremely fast,
even for large values of $\lambda $, for which one would not expect such a
behaviour. This can be explained as follows: In the domain of unbroken $%
\mathcal{PT}$-symmetry $U_{\mathcal{PT}}$ the allowed values for $\kappa $
become very small as $\lambda $ increases. As we note from the expressions (%
\ref{hhh11})-(\ref{hhh14}) the order of $\kappa $ increases with the order
of $\lambda $ term by term.

\subsection{The $N=3$ case}

We will now carry out an analogous perturbative study in $\kappa $ for the
three sites case. We keep the choice of periodic boundary condition, even
though for sites more than two this means some loss of generality.
Proceeding as before, we will try to obtain the matrix $q$ perturbatively,
by solving the consistency conditions (\ref{c1})-(\ref{c5}). Now we have to
solve the problem for $8\times 8$-matrices. We commence by computing the
kernel of $h_{0}$%
\begin{equation*}
\begin{array}{llll}
B_{1}=\mathbb{I}, & B_{2}=S_{zz}^{3}-{\lambda }S_{yyz}^{3}, & B_{3}=\lambda
S_{yy}^{3}-(1-\lambda ^{2})S_{yyz}^{3}-S_{xxz}^{3}, & 
B_{4}=S_{xy}^{3}-S_{yx}^{3}, \\ 
B_{5}=S_{zzz}^{3}, & B_{6}=S_{xyz}^{3}-S_{yxz}^{3}, & B_{7}=\lambda
S_{xx}^{3}+S_{z}^{3}=-2h_{0}(\lambda ), & B_{8}=S_{xx}^{3}+S_{yy}^{3}+%
\lambda S_{yyz}^{3},%
\end{array}%
\end{equation*}%
in addition to this eight matrices, there are another four, due to the fact
that two of the eigenvalues of $h_{0}(\lambda )$ are degenerate. Hence the
dimension of the kernel is 12, 
\begin{eqnarray}
B_{9} &=&S_{z}^{3}-\lambda (S_{yy}^{3}+S_{zz}^{3}-\sigma _{1}^{y}\sigma
_{3}^{y}-\sigma _{1}^{z}\sigma _{3}^{z}-\sigma _{1}^{x}\sigma
_{3}^{x}),\quad B_{10}=\sigma _{2}^{y}\sigma _{3}^{y}+\sigma _{2}^{z}\sigma
_{3}^{z}+\sigma _{2}^{x}\sigma _{3}^{x}, \\
B_{11} &=&S_{zz}^{3}+\lambda S_{xxz}^{3}-\lambda (\sigma _{1}^{z}+\sigma
_{3}^{z}+\sigma _{1}^{x}\sigma _{2}^{z}\sigma _{3}^{x}+\sigma _{1}^{y}\sigma
_{2}^{z}\sigma _{3}^{y}),\quad B_{12}=\sigma _{3}^{z}-\sigma _{1}^{x}\sigma
_{2}^{x}\sigma _{3}^{z}-\sigma _{1}^{y}\sigma _{2}^{y}\sigma _{3}^{z}, 
\notag
\end{eqnarray}%
with $[B_{i},h_{0}(\lambda )]=0$ for $i=1,\ldots ,12$. Similarly as in the
case $N=2$ we find that all of these matrices are parity invariant 
\begin{equation}
\mathcal{P}B_{i}\mathcal{P}=B_{i},\quad \forall \quad i=1,\ldots ,8,
\label{all}
\end{equation}%
which from equations (\ref{cons}) means that no linear combination of the
matrices $B_{i}$ can be added to $q_{2k-1}$ that would be compatible with
the constraints (\ref{ecu}). Therefore, with such constraints, there is a
unique solution to (\ref{c1}) which has the form, 
\begin{equation}
q_{1}(\lambda )=-S_{y}^{3}-\lambda (S_{yz}^{3}+S_{zy}^{3})+2\lambda
^{2}(S_{yyy}^{3}-S_{zzy}^{3}).
\end{equation}%
As we can see, the two first terms in $q_{1}(\lambda )$ are a direct
generalization of the result for two sites, which hints at the existence of
a general pattern. As for the $N=2$ case we find once again, that even
before attempting to solve (\ref{c1}), we could have predicted from (\ref%
{cons}) that the matrices $q_{2k-1}(\lambda )$ can only be linear
combinations of $S_{y}^{3},S_{yz}^{3},S_{zy}^{3},S_{yyy}^{3},S_{zzy}^{3}$
and $S_{xxy}^{3}$ (for $k=1$, equation (\ref{BB}) tells us though that the
coefficient of $S_{xxy}^{3}$ is zero. This will change for higher orders in
perturbation theory). We can therefore write, 
\begin{equation}
q=\hat{\alpha}(\lambda ,\kappa )S_{y}^{3}+\hat{\beta}(\lambda ,\kappa
)(S_{yz}^{3}+S_{zy}^{3})+\hat{\gamma}(\lambda ,\kappa )S_{yyy}^{3}+\hat{%
\delta}(\lambda ,\kappa )S_{xxy}^{3}+\hat{\epsilon}(\lambda ,\kappa
)S_{zzy}^{3},
\end{equation}%
where 
\begin{eqnarray}
\hat{\alpha}(\lambda ,\kappa ) &=&\sum\nolimits_{k=1}^{\infty }\hat{a}%
_{2k-1}(\lambda )\kappa ^{2k-1},\quad \hat{\beta}(\lambda ,\kappa
)=\sum\nolimits_{k=1}^{\infty }\hat{b}_{2k-1}(\lambda )\kappa ^{2k-1},
\label{al} \\
\hat{\gamma}(\lambda ,\kappa ) &=&\sum\nolimits_{k=1}^{\infty }\hat{s}%
_{2k-1}(\lambda )\kappa ^{2k-1},\quad \hat{\delta}(\lambda ,\kappa
)=\sum\nolimits_{k=1}^{\infty }\hat{d}_{2k-1}(\lambda )\kappa ^{2k-1}, \\
\hat{\epsilon}(\lambda ,\kappa ) &=&\sum\nolimits_{k=1}^{\infty }\hat{e}%
_{2k-1}(\lambda )\kappa ^{2k-1}.
\end{eqnarray}%
Computing coefficients up to order $\kappa ^{7}$ we find the results in
tables 5-7.

\begin{center}
{\small 
\begin{tabular}{|c||c|c|c|c|c|c|c|}
\hline
& $-\lambda^0 $ & $-\lambda^2$ & $-\lambda^4$ & $-\lambda^6$ & $-\lambda^8$
& $-\lambda^{10}$ & $-\lambda^{12}$ \\ \hline\hline
$\hat{a}_1(\lambda) $ & 1 & 0 & 0 & 0 & 0 & 0 & 0 \\ \hline
$\hat{a}_3(\lambda) $ & $\frac{1}{3}$ & $\frac{8}{3}$ & 16 & 0 & 0 & 0 & 0
\\ \hline
$\hat{a}_5(\lambda) $ & $\frac{1}{5}$ & $\frac{122}{15}$ & 144 & $\frac{5024%
}{15} $ & $\frac{3072}{5}$ & 0 & 0 \\ \hline
$\hat{a}_7(\lambda) $ & $\frac{1}{7}$ & $\frac{576}{35}$ & $\frac{9616}{15} $
& $\frac{432832}{105}$ & $\frac{1755136}{105}$ & $\frac{2720768}{105}$ & $%
\frac{196608}{7}$ \\ \hline
$\hat{d}_1(\lambda) $ & 0 & 0 & 0 & 0 & 0 & 0 & 0 \\ \hline
$\hat{d}_3(\lambda) $ & 0 & 0 & $\frac{2^4}{3}$ & 0 & 0 & 0 & 0 \\ \hline
$\hat{d}_5(\lambda) $ & 0 & $\frac{2}{3}$ & $\frac{496}{15}$ & $\frac{1184}{%
15} $ & $\frac{2^{10}}{5}$ & 0 & 0 \\ \hline
$\hat{d}_7(\lambda) $ & 0 & $\frac{2^5}{15}$ & $\frac{4432}{35}$ & $\frac{%
86848}{105}$ & $\frac{65024}{15}$ & $\frac{754688}{105}$ & $\frac{2^{16}}{7}$
\\ \hline
\end{tabular}
}
\end{center}

\noindent {\small {\textbf{Table 5:} The coefficients $\hat{a}%
_{2k+1}(\lambda )$ and $\hat{d}_{2k+1}(\lambda )$ for $k<4$.}}

\begin{center}
{\small 
\begin{tabular}{|c||c|c|c|c|c|c|c|}
\hline
& $-\lambda$ & $-\lambda^3$ & $-\lambda^5$ & $-\lambda^7$ & $-\lambda^{9}$ & 
$-\lambda^{11}$ & $-\lambda^{13} $ \\ \hline\hline
$\hat{b}_1(\lambda)$ & 1 & 0 & 0 & 0 & 0 & 0 & 0, \\ 
$\hat{b}_3(\lambda)$ & $\frac{4}{3}$ & $\frac{28}{3}$ & $\frac{2^6}{3}$ & 0
& 0 & 0 & 0 \\ 
$\hat{b}_5(\lambda)$ & $\frac{23}{15}$ & $\frac{664}{15}$ & $\frac{1568}{5}$
& $\frac{3328}{5}$ & $\frac{2^{12}}{5}$ & 0 & 0, \\ 
$\hat{b}_7(\lambda)$ & $\frac{176}{105}$ & $\frac{4344}{35}$ & $\frac{13536}{%
7}$ & $\frac{52416}{5}$ & $\frac{1104384}{35}$ & $\frac{311296}{7}$ & $\frac{%
2^{18}}{7}$ \\ \hline
\end{tabular}
}
\end{center}

\noindent {\small {\textbf{Table 6:} The coefficients $\hat{b}%
_{2k+1}(\lambda )$ for $k<4$.}}

\begin{center}
{\small 
\begin{tabular}{|c||c|c|c|c|c|c|c|}
\hline
& $-\lambda^2$ & $-\lambda^4$ & $-\lambda^6$ & $-\lambda^8$ & $-\lambda^{10}$
& $-\lambda^{12}$ & $-\lambda^{14} $ \\ \hline\hline
$\hat{s}_1(\lambda) $ & -2 & 0 & 0 & 0 & 0 & 0 & 0 \\ \hline
$\hat{s}_3(\lambda) $ & -4 & -8 & $-\frac{2^7}{3}$ & 0 & 0 & 0 & 0 \\ \hline
$\hat{s}_5(\lambda) $ & $-\frac{28}{5}$ & $-\frac{112}{5}$ & $-\frac{2592}{5}
$ & $-\frac{4608}{5}$ & $-\frac{2^{13}}{5}$ & 0 & 0 \\ \hline
$\hat{s}_7(\lambda)$ & $-\frac{232}{35}$ & $\frac{288}{35}$ & $- \frac{91008%
}{35}$ & $- \frac{452224}{35}$ & $-\frac{356352}{7}$ & $- \frac{491520}{7}$
& $\frac{2^{19}}{7}$ \\ \hline
$\hat{e}_1(\lambda) $ & 2 & 0 & 0 & 0 & 0 & 0 & 0 \\ \hline
$\hat{e}_3(\lambda) $ & $\frac{20}{3}$ & $\frac{56}{3}$ & $\frac{2^7}{3}$ & 0
& 0 & 0 & 0 \\ \hline
$\hat{e}_5(\lambda) $ & $\frac{196}{15} $ & $\frac{400}{3}$ & $\frac{3872}{5}
$ & $\frac{6656}{5}$ & $\frac{2^{13}}{5}$ & 0 & 0 \\ \hline
$\hat{e}_7(\lambda)$ & $\frac{440}{21}$ & $\frac{53152}{105}$ & $\frac{206336%
}{35}$ & $\frac{75904}{3}$ & ${69632}$ & $\frac{622592}{7}$ & $\frac{2^{19}}{%
7}$ \\ \hline
\end{tabular}
}
\end{center}

\noindent {\small {\textbf{Table 7:} The coefficients $\hat{s}%
_{2k+1}(\lambda )$ and $\hat{e}_{2k+1}(\lambda )$ for $k<4$.}}

It is now possible to use these perturbative results to compute $h(\lambda
,\kappa )$ for particular values of $\lambda $ and $\kappa $. We find that
the structure of the Hermitian counterpart of the original Hamiltonian is: 
\begin{eqnarray}
h(\lambda ,\kappa ) &=&\mu _{xx}^{3}(\lambda ,\kappa )S_{xx}^{3}+\mu
_{yy}^{3}(\lambda ,\kappa )S_{yy}^{3}+\mu _{zz}^{3}(\lambda ,\kappa
)S_{zz}^{3}+\mu _{z}^{3}(\lambda ,\kappa )S_{z}^{3}  \notag \\
&&+\mu _{xxz}^{3}(\lambda ,\kappa )S_{xxz}^{3}+\mu _{yyz}^{3}(\lambda
,\kappa )S_{yyz}^{3}+\mu _{zzz}^{3}(\lambda ,\kappa )S_{zzz}^{3},
\end{eqnarray}%
which resembles the result for two sites, but includes few extra terms that
couple all three sites. The functions $\mu _{xx}^{3},\ldots ,\mu _{zzz}^{3}$
are all real functions of the couplings. As for $N=2$, the Hamiltonian above
is $\mathcal{PT}$-symmetric, which follows from the fact that all matrices
involved are invariant under the adjoint action of the operator $\mathcal{PT}
$ (see equation (\ref{ptmat})). As for $N=2$ also, these are the only
matrices that are both $\mathcal{PT}$ symmetric and real (notice that, from
the definition (\ref{matrices}) for $N=3$, it holds that $%
S_{xxz}^{3}=S_{zxx}^{3}=S_{xzx}^{3}$ and $%
S_{yyz}^{3}=S_{zyy}^{3}=S_{yzy}^{3} $).

\subsection{The $N=4$ case}

It is interesting to investigate how the perturbative results generalize as
we increase the number of sites. The $N=4$ case is especially interesting as
it is the simplest example for which we may see non local interaction terms
in the Hermitian Hamiltonian. There is again only one solution for $%
q_{1}(\lambda )$ which is compatible with the conditions (\ref{cons}), that
is 
\begin{eqnarray}
q_{1}(\lambda ) &=&-S_{y}^{4}-\lambda (S_{yz}^{4}+S_{zy}^{4})-\frac{6\lambda
^{3}(S_{yuz}^{4}-S_{yz}^{4}-S_{zy}^{4})}{40\lambda ^{2}-9}  \notag \\
&&+\frac{1}{40\lambda ^{2}-9}\left[ (9-32\lambda ^{2})\lambda
^{2}(S_{yzz}^{4}+S_{zzy}^{4})-32\lambda ^{4}S_{zyz}^{4}-2\lambda
^{2}(3-16\lambda ^{2})S_{yyy}^{4}\right.  \notag \\
&&-\left. 3\lambda ^{2}(S_{xxy}^{4}-2S_{xyx}^{4}+S_{yxx}^{4})+2\lambda
^{3}(S_{xxyz}^{4}-5S_{xyxz}^{4}+S_{xxzy}^{4})\right.  \notag \\
&&+\left. 2\lambda ^{3}(9S_{yzzz}^{4}-7S_{yyyz}^{4})+64\lambda
^{5}(S_{yyyz}^{4}-S_{zzzy}^{4})\right] .  \label{for}
\end{eqnarray}%
In many ways, this is a simple generalization of the results of two and
three sites. The matrices that enter the expression are to a large extent
the same we find for less sites, but we have now extra contributions
involving Pauli matrices sitting at all four sites of the chain, which was
to be expected. There are however two major changes

\begin{itemize}
\item the dependence on $\lambda $ of the coefficients is not polynomial
anymore,

\item the first occurrence of non-local interactions appears through the
matrix $S_{yuz}^{4}$.
\end{itemize}

As for lower values of $N$, it is not difficult to argue that the matrices (%
\ref{matrices}) entering the linear combination (\ref{for}) are the only
ones that are compatible with (\ref{ecu}). Hence, as expected, the same
structure extends to higher orders in perturbation theory, although
expressions become extremely involved. The table below gives $q_{3}(\lambda
) $ as a sum of terms given by the matrices on the first column multiplied
by the corresponding coefficients in the second column,

\begin{center}
{\small 
\begin{tabular}{|l|c|}
\hline
$q_{3}(\lambda )$ & \text{Coefficients} \\ \hline
$S_{y}^{4}$ & $\frac{-81+72{\lambda }^{2}+1892{\lambda }^{4}-4224{\lambda }%
^{6}+28672{\lambda }^{8}-131072{\lambda }^{10}}{3{\left( -9+40{\lambda }%
^{2}\right) }^{2}}$ \\ \hline
$S_{yz}^{4}+S_{zy}^{4}$ & $\frac{2916\lambda -22842{\lambda }^{3}+27216{%
\lambda }^{5}+81152{\lambda }^{7}+251904{\lambda }^{9}+786432{\lambda }%
^{11}-6291456{\lambda }^{13}}{3{\left( -9+40{\lambda }^{2}\right) }^{3}}$ \\ 
\hline
$S_{yuz}^{4}$ & $-\frac{64{\lambda }^{5}\left( 351+276{\lambda }^{2}-8352{%
\lambda }^{4}-4096{\lambda }^{6}+65536{\lambda }^{8}\right) }{3{\left( -9+40{%
\lambda }^{2}\right) }^{3}}$ \\ \hline
$S_{yzz}^{4}+S_{zzy}^{4}$ & $-\frac{4{\lambda }^{2}\left( -1215+7722{\lambda 
}^{2}+12432{\lambda }^{4}-151808{\lambda }^{6}+131072{\lambda }^{8}-262144{%
\lambda }^{10}+2097152{\lambda }^{12}\right) }{3{\left( -9+40{\lambda }%
^{2}\right) }^{3}}$ \\ \hline
$S_{zyz}^{4}$ & $\frac{4{\lambda }^{2}\left( 1215-16065{\lambda }^{2}+26952{%
\lambda }^{4}+72448{\lambda }^{6}-2097152{\lambda }^{12}\right) }{3{\left(
-9+40{\lambda }^{2}\right) }^{3}}$ \\ \hline
$S_{yyy}^{4}$ & $\frac{4{\lambda }^{2}\left( -729+8667{\lambda }^{2}-14040{%
\lambda }^{4}-97024{\lambda }^{6}+393216{\lambda }^{8}-1048576{\lambda }%
^{10}+2097152{\lambda }^{12}\right) }{3{\left( -9+40{\lambda }^{2}\right) }%
^{3}}$ \\ \hline
$S_{xxy}^{4}+S_{yxx}^{4}$ & $-\frac{8{\lambda }^{2}\left( 243-2079{\lambda }%
^{2}+7752{\lambda }^{4}+16768{\lambda }^{6}-159744{\lambda }^{8}+262144{%
\lambda }^{10}\right) }{3{\left( -9+40{\lambda }^{2}\right) }^{3}}$ \\ \hline
$S_{xyx}^{4}$ & $-\frac{4{\lambda }^{2}\left( -972+6939{\lambda }^{2}-31512{%
\lambda }^{4}+82176{\lambda }^{6}-188416{\lambda }^{8}+262144{\lambda }%
^{10}\right) }{3{\left( -9+40{\lambda }^{2}\right) }^{3}}$ \\ \hline
$S_{xxyz}^{4}+S_{xxzy}^{4}$ & $-\frac{2{\lambda }^{3}\left( 405-19368{%
\lambda }^{2}+146048{\lambda }^{4}-349184{\lambda }^{6}-131072{\lambda }%
^{8}+1048576{\lambda }^{10}\right) }{3{\left( -9+40{\lambda }^{2}\right) }%
^{3}}$ \\ \hline
$S_{xyxz}^{4}$ & $-\frac{64{\lambda }^{3}\left( 81-99{\lambda }^{2}-692{%
\lambda }^{4}+672{\lambda }^{6}+4096{\lambda }^{8}\right) }{3{\left( -9+40{%
\lambda }^{2}\right) }^{3}}$ \\ \hline
$S_{zyyy}^{4}$ & $\frac{32{\lambda }^{3}\left( -567+6390{\lambda }^{2}-21448{%
\lambda }^{4}+12096{\lambda }^{6}+49152{\lambda }^{8}-196608{\lambda }%
^{10}+524288{\lambda }^{12}\right) }{3{\left( -9+40{\lambda }^{2}\right) }%
^{3}}$ \\ \hline
$S_{yzzz}^{4}$ & $-\frac{32{\lambda }^{3}\left( -729+7290{\lambda }^{2}-19512%
{\lambda }^{4}-5952{\lambda }^{6}+32768{\lambda }^{8}-65536{\lambda }%
^{10}+524288{\lambda }^{12}\right) }{3{\left( -9+40{\lambda }^{2}\right) }%
^{3}}$ \\ \hline
\end{tabular}
}
\end{center}

\noindent In general we have, 
\begin{eqnarray}
&&q=\zeta (\lambda ,\kappa )S_{y}^{4}+\theta (\lambda ,\kappa
)(S_{zy}^{4}+S_{yz}^{4})+\vartheta (\lambda ,\kappa )S_{yuz}^{4}  \notag
\label{akg} \\
&&+\mu (\lambda ,\kappa )(S_{yzz}^{4}+S_{zzy}^{4})+\nu (\lambda ,\kappa
)S_{zyz}^{4}+\xi (\lambda ,\kappa )S_{yyy}^{4}+\varpi (\lambda ,\kappa
)(S_{xxy}^{4}+S_{yxx}^{4}) \\
&&+\varrho (\lambda ,\kappa )S_{xyx}^{4}+\varsigma (\lambda ,\kappa
)(S_{xxyz}^{4}+S_{xxzy}^{4})+\tau (\lambda ,\kappa )S_{xyxz}^{4}+\upsilon
(\lambda ,\kappa )S_{zyyy}^{4}+\chi (\lambda ,\kappa )S_{yzzz}^{4},  \notag
\end{eqnarray}%
where all coefficients $\zeta (\lambda ,\kappa ),\theta (\lambda ,\kappa
),\ldots ,\chi (\lambda ,\kappa )$ can be expressed as expansions of the
form (\ref{al}) and are real functions of the couplings. Perturbation theory
results show that the Hermitian Hamiltonian $h(\lambda ,\kappa )$ has the
following structure: 
\begin{eqnarray}
h(\lambda ,\kappa ) &=&\mu _{xx}^{4}(\lambda ,\kappa )S_{xx}^{4}+\nu
_{xx}^{4}(\lambda ,\kappa )S_{xux}^{4}+\mu _{yy}^{4}(\lambda ,\kappa
)S_{yy}^{4}+\nu _{yy}^{4}(\lambda ,\kappa )S_{yuy}^{4}  \notag \\
&&+\mu _{zz}^{4}(\lambda ,\kappa )S_{zz}^{4}+\nu _{zz}^{4}(\lambda ,\kappa
)S_{zuz}^{4}+\mu _{z}^{4}(\lambda ,\kappa )S_{z}^{4}+\mu _{xxz}^{4}(\lambda
,\kappa )(S_{xxz}^{4}+S_{zxx}^{4})  \notag \\
&&+\mu _{xzx}^{4}(\lambda ,\kappa )S_{xzx}^{4}+\mu _{yyz}^{4}(\lambda
,\kappa )(S_{yyz}^{4}+S_{zyy}^{4})+\mu _{yzy}^{4}(\lambda ,\kappa
)S_{yzy}^{4}+\mu _{zzz}^{4}(\lambda ,\kappa )S_{zzz}^{4}  \notag \\
&&+\mu _{xxxx}^{4}(\lambda ,\kappa )S_{xxxx}^{4}+\mu _{yyyy}^{4}(\lambda
,\kappa )S_{yyyy}^{4}+\mu _{zzzz}^{4}(\lambda ,\kappa )S_{zzzz}^{4}+\mu
_{xxyy}^{4}(\lambda ,\kappa )S_{xxyy}^{4}  \notag \\
&&+\mu _{xyxy}^{4}(\lambda ,\kappa )S_{xyxy}^{4}+\mu _{zzyy}^{4}(\lambda
,\kappa )S_{zzyy}^{4}+\mu _{zyzy}^{4}(\lambda ,\kappa )S_{zyzy}^{4}+\mu
_{xxzz}^{4}(\lambda ,\kappa )S_{xxzz}^{4}  \notag \\
&&+\mu _{xzxz}^{4}(\lambda ,\kappa )S_{xzxz}^{4}.  \label{h4lk}
\end{eqnarray}%
As expected from the expression of $q$, we find that $h(\lambda ,\kappa )$
involves non-local interaction terms proportional to $S_{xux}^{4}$, $%
S_{yuy}^{4}$ and $S_{zuz}^{4}$. The remaining terms are the natural
generalization of the those appearing for the $N=2,3$ cases plus additional
terms corresponding to interactions that couple all four sites of the chain.
Once again, all coefficients $\mu _{xx}^{4},\ldots ,\mu _{xzxz}^{4}$ are
real functions of the couplings. As for previous cases, it turns out that
matrices appearing in the linear combination (\ref{h4lk}) are exactly those
that are both invariant under $\mathcal{PT}$-symmetry, according to equation
(\ref{ptmat}), and real.

\subsection{Some general features from perturbation theory}

\label{pertu}

We would like to end this section by summarizing the main results that we
have obtained from our perturbative analysis. Since we have only solved for $%
2,3$ and 4 sites, our conclusions are based on a case-by-case analysis
rather than rigorous proofs. However, we believe that the consistent
occurrence of certain features across the various examples that we have
studied provides strong support for these conclusions.

Firstly we found that the combination of perturbation theory and the
assumption of Hermiticity of the Dyson operator $\eta =e^{q/2}$ fix the
metric $\rho $ and therefore the Hermitian Hamiltonian $h(\lambda ,\kappa )$
with its corresponding observables \textit{uniquely}. We have established
this for $N=2,3,4$ and arbitrary values of both coupling constants as well
as for arbitrary $N$ if $\lambda =0$.

Secondly, concerning the specific algebraic structure of the Hermitian
Hamiltonian, we have seen that it becomes more involved for higher values of 
$N$. For $N>2$ it generally includes interaction terms that couple two or
more adjacent sites, as well as non-local terms that couple non-adjacent
sites. In addition, this structure is entirely dictated by $\mathcal{PT}$
symmetry, which selects out which tensor products of Pauli and identity
matrices the Hamiltonian will be a linear combination of. Combining the
requirement of $\mathcal{PT}$ symmetry with the requirement of $h(\lambda
,\kappa )$ being real completely fixes the general structure of $h(\lambda
,\kappa )$, although not the specific dependence on the coupling constants $%
\lambda $ and $\kappa $, which is fixed by perturbation theory. All examples
studied indicate that for a given value of $N$, all solutions $%
q_{2k-1}(\lambda )$, with $k \geq 0$ at different perturbative orders, share
a common structure, namely they are all linear combinations of the same set
of matrices, with coefficients that increase in complexity with increasing
values of $k$.

Finally, concerning the numerical accuracy of perturbation theory, we have
demonstrated in detail that it converges very quickly for $N=2$. For $N=2,3$
and 4 it becomes very difficult to perform computations up to such high
orders of perturbation theory reached for $N=2$ and the rate of convergence
has not been analysed in detail for such cases. An interesting aspect of the
model studied here is the dependence of the Hamiltonian on two coupling
constants. For $N=2$, we have carried out perturbation theory in both such
couplings and found quick convergence in both cases. All our perturbation
theory results, suggest that the entries of the Hermitian Hamiltonian $%
h(\lambda ,\kappa )$ can generally be expressed as a double Taylor series in 
$\lambda $ and $\kappa $.

\section{Expectation values of local operators: form factors}

In this section we want to employ our general formulae in order to compute
the expectation values of certain local operators of the chain. In
particular, we will be looking at the expectation values of the total spin
in the $x$ and $z$ directions in the ground state of the chain. These
expectation values are commonly known as the magnetization in the $x$ and $z$
directions. Recalling the results from section \ref{local}, we define 
\begin{eqnarray}
M_{z}(\lambda ,\kappa ) &=&\frac{1}{2}\langle \Psi _{g}|\eta S_{z}^{N}\eta
|\Psi _{g}\rangle =\frac{1}{2}\langle \psi _{g}|S_{z}^{N}|\psi _{g}\rangle ,
\label{mag1} \\
M_{x}(\lambda ,\kappa ) &=&\frac{1}{2}\langle \Psi _{g}|\eta S_{x}^{N}\eta
|\Psi _{g}\rangle =\frac{1}{2}\langle \psi _{g}|S_{x}^{N}|\psi _{g}\rangle .
\label{mag}
\end{eqnarray}
where $|\psi _{g}\rangle $ is the ground state of the Hermitian Hamiltonian
and $|\Psi _{g}\rangle $ is the ground state of the non-Hermitian one. We
assume that the states are normalized to $\langle \psi _{g}|\psi _{g}\rangle
=\langle \Psi _{g}|\Psi _{g}\rangle =1$. In the following sections, we will
carry out this computation for $\lambda =0$ with generic $N$ and for $%
\lambda \neq 0$ for small values of $N$.

\subsection{General solutions for $\protect\lambda=0$}

\noindent In section \ref{known} we described in detail how for $\lambda =0$
the original Hamiltonian and its Hermitian counterpart simplify greatly.
Indeed, the latter can be found in all generality, for any number of sites,
resulting in the expression (\ref{final}). Taking (\ref{diag}) and (\ref{gss}%
) into account, it is very easy to show that 
\begin{equation}
M_{z}(0,\kappa )=\frac{N}{2},  \label{max}
\end{equation}
which is nothing but the total spin of the chain and does not depend on the
particular value of the coupling $\kappa $. This result is to be expected
for a Hamiltonian like (\ref{final}). Naturally, the spins of the chain tend
to align in the direction of the field, and will all be up so that the
magnetization is just the total spin of the chain and maximal. A similar
computation can be performed for $M_{x}(0,\kappa )$ for each particular
value of $N$. In all cases one finds 
\begin{equation}
M_{x}(0,\kappa )=0,
\end{equation}
which is also what one would expect for this model, as the Hamiltonian (\ref%
{final}) does not favour any particular direction of the spin $\sigma _{x}$.

\subsection{General solutions for $\protect\kappa=0$}

\noindent For $\kappa =0$ the Hamiltonian (\ref{H}) is Hermitian and
therefore computations of the magnetization simplify, as $\eta =\mathbb{I}$.
The ground state will nonetheless still depend on the value of $\lambda $.
For example, for $N=2$ it is 
\begin{equation}
|\psi _{g}\rangle =\frac{1}{\sqrt{2(1+{\lambda }^{2}+\sqrt{1+{\lambda }^{2}})%
}}\left( 
\begin{array}{c}
{1+\sqrt{1+{\lambda }^{2}}} \\ 
0 \\ 
0 \\ 
{\lambda }%
\end{array}
\right)  \label{gs}
\end{equation}
with energy $E_{g}=-\sqrt{1+\lambda ^{2}}$ and the magnetizations becomes
simply 
\begin{equation}
M_{z}(\lambda ,0)=\frac{1}{\sqrt{1+\lambda ^{2}}}\qquad \text{and\qquad }%
M_{x}(\lambda ,0)=0.  \label{mag0}
\end{equation}
The function $M_{z}(\lambda ,0)$ flows between the value $M_{z}(0,0)=1$, as
seen in the previous section, and $M_{z}(\infty ,0)\rightarrow 0$. This is
simply because for $\kappa =0$ our model is nothing but the Ising chain with
a magnetic field of intensity $1/\lambda $ in the $z$-direction. Therefore,
as $\lambda \rightarrow \infty $ the intensity of the perturbing field tends
to zero, and the ground state of the chain has zero magnetization, as
consecutive spins align in opposite directions to minimize energy. This is a
general feature that will also hold for higher values of $N$. For example,
we find 
\begin{eqnarray}
\mu (\lambda ,0) &=&\frac{1}{2}+\frac{2-\lambda }{2\sqrt{1+\left( -1+\lambda
\right) \lambda }}\quad \text{for $N=3$},  \label{mu3} \\
\mu (\lambda ,0) &=&\frac{\left( 1-{\lambda }^{2}+\sqrt{1+{\lambda }^{4}}%
\right) \sqrt{1+{\lambda }^{2}+\sqrt{1+{\lambda }^{4}}}}{\sqrt{2}\sqrt{1+{%
\lambda }^{4}}}\quad \text{for $N=4$}.
\end{eqnarray}
In both cases we recover the expected behaviour: $\mu (0,0)=N/2$ and $\mu
(\infty ,0)=0$. The $\kappa =0$ curve in figure 4 is precisely a plot of the
function (\ref{mag0}) for $N=2$.

The second equation in (\ref{mag0}) can also be explained easily as a
consequence of the symmetry of the Hamiltonian $H(\lambda ,0)$. Such
Hamiltonian is invariant under the transformation $\sigma
_{i}^{x}\rightarrow -\sigma _{i}^{x}$ at each site $i$ of the chain. This
means that any form factor involving the operators $\sigma _{i}^{x}$ must
have the same symmetry. Therefore, 
\begin{equation}
M_{x}(\lambda ,0)=-M_{x}(\lambda ,0),  \label{sim}
\end{equation}%
which implies $M_{x}(\lambda ,0)=0$ for all values of $N$.

\subsection{ The $N=2$ case for $\protect\kappa, \protect\lambda \neq 0$}

\noindent Let us now compute $M_z(\lambda,\kappa)$ and $M_x(\lambda,\kappa)$
in the more generic situation when both coupling $\lambda$ and $\kappa$ are
non vanishing. We will start by analyzing the magnetization in the $z$%
-direction. In this case ($\lambda \neq 0$), the form of the ground state of
the Hermitian chain is not particularly simple and therefore we will work
with the first equality in (\ref{mag}) and employ the properly normalized
ground state of the non-Hermitian Hamiltonian. As figure 3 shows, the
magnetization is maximal at $\lambda=0$ with value 1, and exhibits different
kinds of behaviour as $\lambda$ increases, depending of the value of $\kappa$
under consideration.

For every fixed value of $\kappa $, the corresponding graph in figure 3
generally only covers a small region of values of $\lambda $. These are
precisely the values that lie in the region $U_{\mathcal{PT}}$ of figure 1,
namely those values for which all eigenvalues of $H(\lambda ,\kappa )$ are
real. As shown in figure 3, the smaller the value of $\kappa $ the larger
this region becomes in $\lambda $. Depending on the value of $\kappa $ the
magnetization exhibits a rich structure: for $\kappa \geq 0.7$ it is a
strictly decreasing function, whereas for $\kappa \leq 0.6$ it has a
minimum. This minimum is located near the critical value of $\lambda $ above
which some eigenvalues of the Hamiltonian become complex, except for $\kappa
=0.6$, where the minimum of the magnetization shifts to a smaller value of $%
\lambda $.

With regard to the magnetization in the $x$-direction we find that it
vanishes for all values of $\lambda$ and $\kappa$. This is so because the
Hermitian counter-part of $H(\lambda,\kappa)$ with $N=2$ has the form (\ref%
{struc}) and therefore the Hamiltonian $h(\lambda,\kappa)$ has the same
symmetry described at the end of the previous section.

\medskip

\includegraphics[width=12cm,height=8cm]{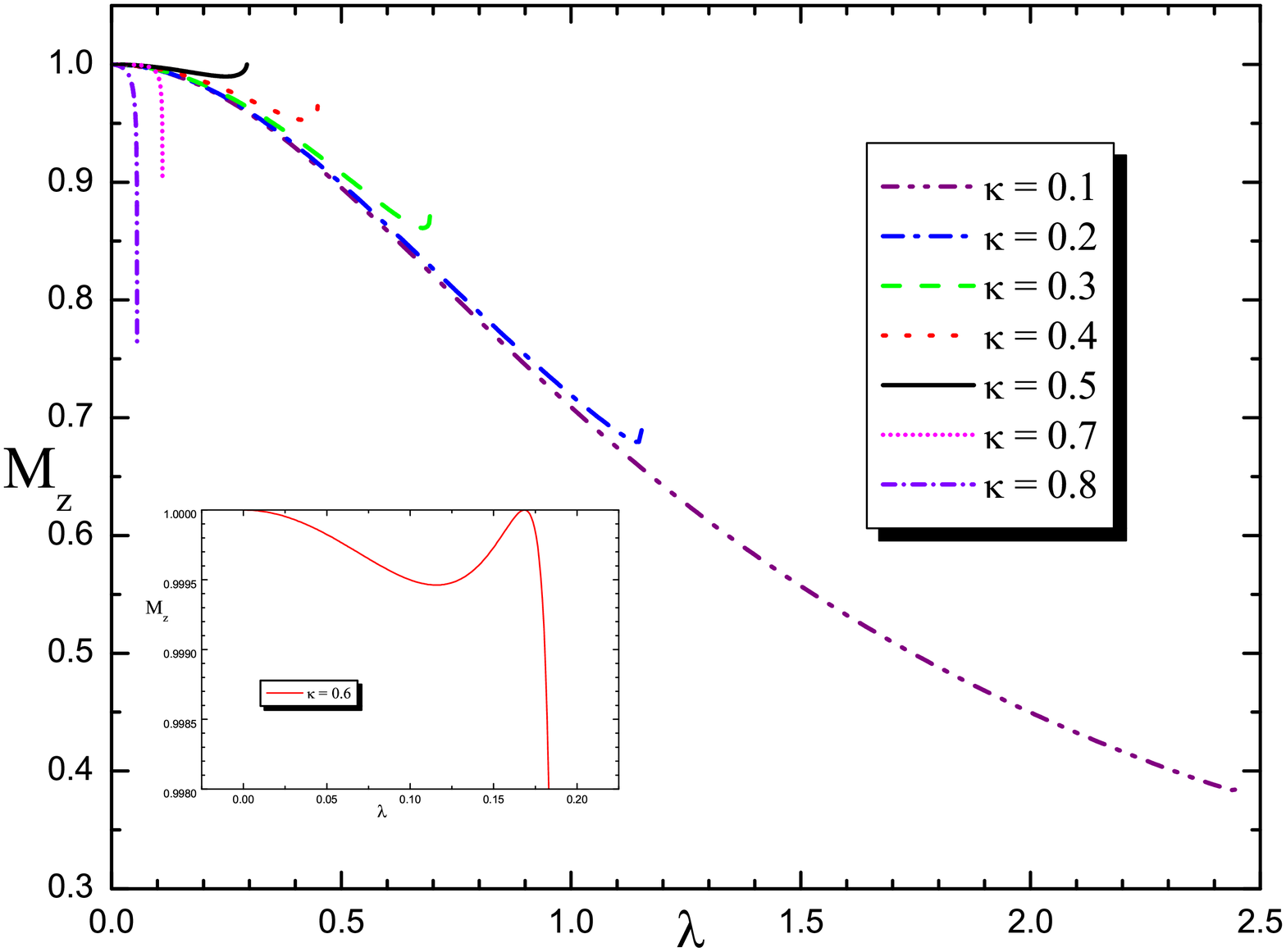}

{\small \noindent {\textbf{Figure 3:} The magnetization in the $z$-direction
for $N=2$ as a function of $\lambda$ and $\kappa$.} }

It is also interesting to analyze how the presence of an imaginary magnetic
field in the $x$-direction in (\ref{H}), as opposed to a real one really
changes the physics of the model. Figure 4 precisely shows the magnitude of
that change for the magnetization when $\kappa $ is an imaginary number. The
Hamiltonian (\ref{H}) is now that of the Ising spin chain with both a
perpendicular and longitudinal fields applied at each site of the chain. The
competition between these two fields will determine the values of the
magnetization in both the $x$ and $z$-directions.

\medskip

\noindent {\small \includegraphics[width=7.5cm,height=5cm]{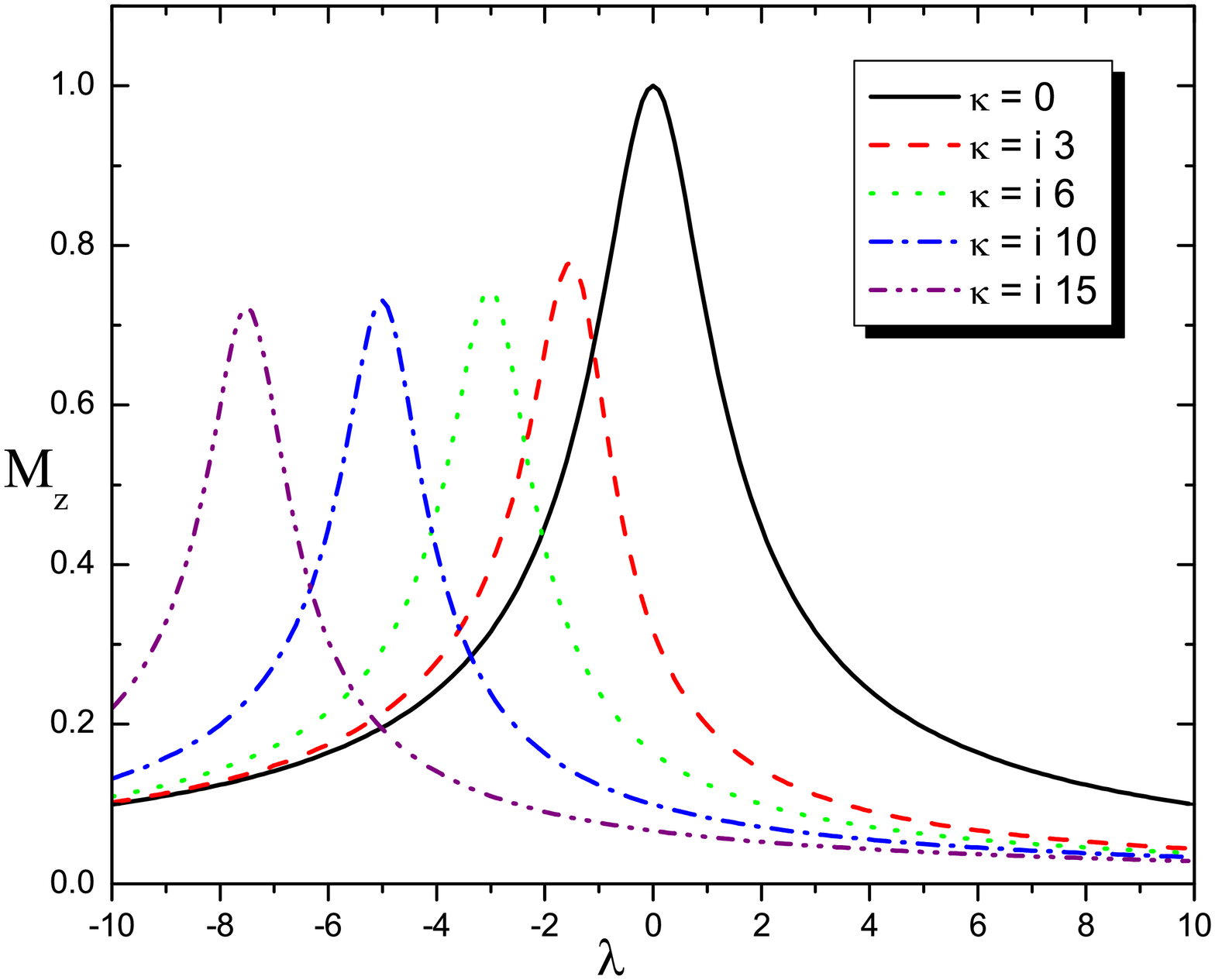} %
\includegraphics[width=7.5cm,height=5cm]{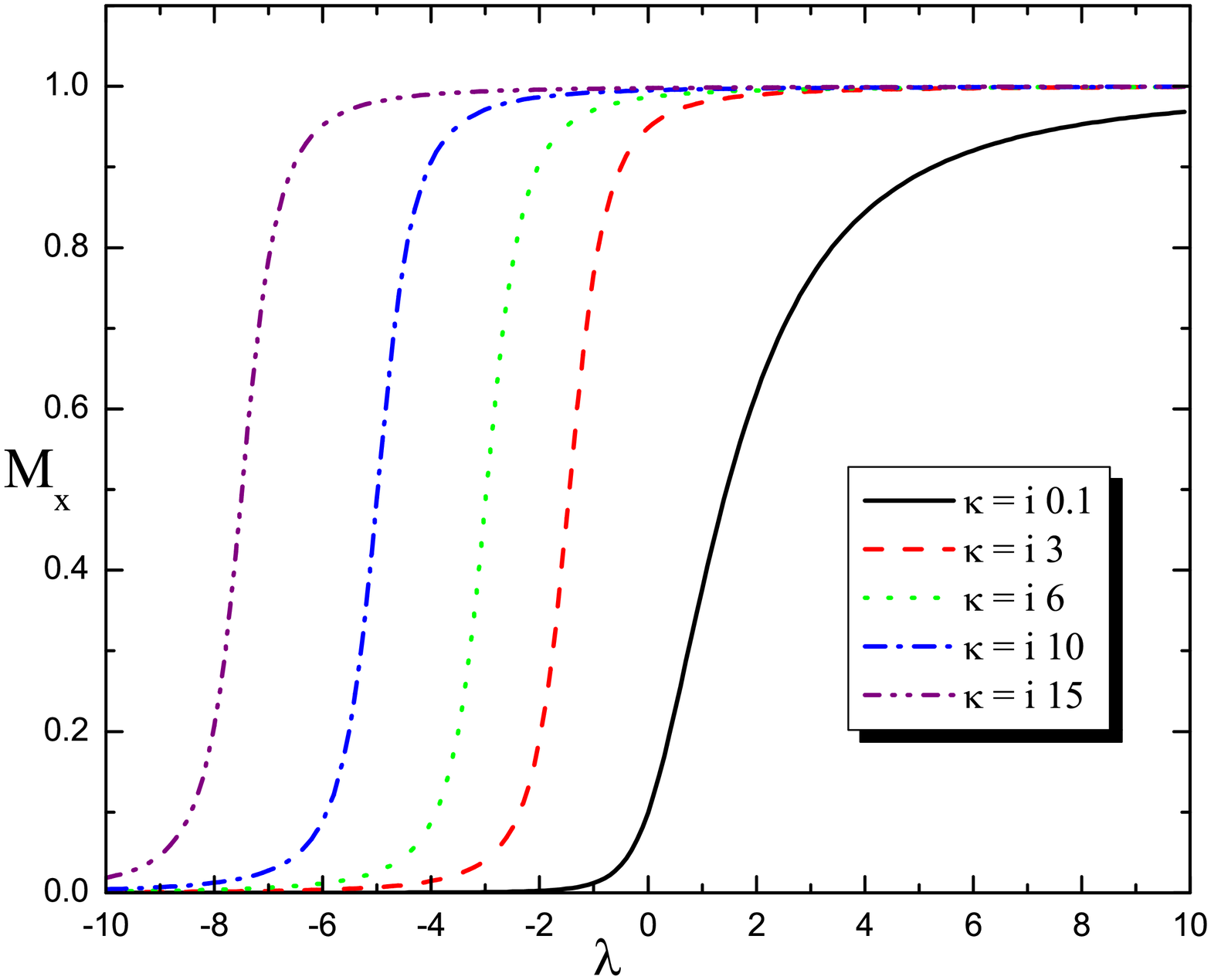} }

{\small \noindent {\textbf{Figure 4:} The magnetization in the $x$ and $z$
directions for $N=2$ and $\kappa $ imaginary.} }

\medskip

We also observe that the magnetization is strictly smaller than 1, as it
should be. Computing the expressions (\ref{mag1}) and (\ref{mag}) in the
standard metric $\rho =\mathbb{I}$, i.e. disregarding the fact that the
Hamiltonian is non-Hermitian, leads to non-physical values larger than one.

\section{Conclusions}

We have demonstrated that there are various possibilities to implement $%
\mathcal{PT}$-symmetry for quantum spin chains, either as a
\textquotedblleft macro-reflection\textquotedblright\ by reflection across
the entire chain or as \textquotedblleft micro-reflection\textquotedblright\
by reflecting at individual sites. These new possibilities constitute
symmetries for the model $H(\lambda ,\kappa )$ in (\ref{H}) we focussed on,
i.e. Ising quantum spin chain in the presence of a magnetic field in the $z$%
-direction as well as a longitudinal imaginary field in the $x$-direction.
However, there are also implications for other Hamiltonians such as $H_{XXZ}$
in (\ref{XXZ}) and $H_{DG}$ in (\ref{DeGo}). Due to the various
possibilities to implement parity the corresponding metric and therefore the
underlying physical model is more ambiguous and it requires further
clarification as to which physical system it describes. Remarkably the
non-Hermitian Hamiltonian $H(\lambda ,\kappa )$ fixes the underlying physics
uniquely under the sole assumption the Dyson map $\eta $ is Hermitian. As
pointed out above this uniqueness is not obtained in general. One might
conjecture that this is due to the finite dimensionality of the Hilbert
space, as opposed to continuous models studied for instance in \cite%
{Bender:2004sa,Mosta}, but our comments on $H_{XXZ}$ and $H_{DG}$ suggest
this is not the case. The explanation lies surely in the different types of
symmetries a Hamiltonian might possess, which is supported by the fact that
two different types of metric operators, say $\rho $ and $\hat{\rho}$, can
always be used to define a new non-unitary symmetry operator $S=\hat{\rho}%
\rho ^{-1}$ \cite{Mostsyme,PEGAAF}.

We have shown that all these possibilities serve to define anti-linear
operators, which can not only be used to explain the reality of the spectra
and identify the corresponding domains in the coupling constants, but can
also be employed to define a consistent quantum mechanical framework.
Regarding the technical feasibility of this programme, we have demonstrated
for two sites that the perturbation theory, in $\kappa $ as well as in $%
\lambda $, converges very fast by comparing it with the exact result. We
took this as encouragement to tackle also three and four sites, albeit up to
not as high orders of perturbation theory. Our perturbative analysis has
allowed us to demonstrate for specific examples that the combination of
perturbation theory and Hermiticity of the Dyson operator are sufficient to
uniquely fix $\eta, \rho$ and $h(\lambda,\kappa)$. In fact, for the model at
hand, the constraint of Hermiticity of $\eta$ appears to be sufficient to
entirely fix the algebraic structure of these quantities, even before any
perturbative analysis is carried out.

Clearly there are various open issues and follow up problems associated to
our investigations. Firstly one may try to complete the analysis for the
Hamiltonian $H(\lambda ,\kappa )$ by carrying out further numerical studies,
perturbative computations for more sites and ultimately obtain a complete
analytic understanding for instance by means of the Bethe ansatz. Special
attention should be given to the values of $\kappa $ and $\lambda$
corresponding to the exceptional points, when the usual analysis is expected
to break down. Secondly one may consider the model for higher spin values as
for instance studied in \cite{gehlen2}. Finally it would be also very
interesting to investigate some other members of the class belonging to the
perturbed $\mathcal{M}_{p,q}$-series of minimal conformal field theories.

\medskip \noindent \textbf{Acknowledgments}: A.F. is grateful to G\"{u}nther
von Gehlen for bringing the papers \cite{gehlen1,gehlen2} to our attention.
O.C.A. would like to thank Benjamin Doyon for helpful discussions and
suggestions. We are grateful to Pijush K. Ghosh for bringing reference \cite{DeGosh} to our attention and Vincent Caudrelier for comments on the
manuscript.

\appendix

\section{Exact Hermitian Hamiltonian for $N=2$}

\label{A} As demonstrated in section \ref{5} perturbation theory, both in $%
\kappa $ and $\lambda $, agrees numerically very well with the exact results
of section \ref{4}. We showed that the Hermitian counterpart to the
Hamiltonian (\ref{H}) for $N=2$ can be obtained by computing 
\begin{equation}
h(\lambda ,\kappa )=e^{q/2}H(\lambda ,\kappa )e^{-q/2},\quad \text{with}%
\quad q=\alpha (\lambda ,\kappa )S_{y}^{2}+\beta (\lambda ,\kappa
)(S_{yz}^{2}+S_{zy}^{2}),  \label{above}
\end{equation}
where the functions $\alpha (\lambda ,\kappa )$ and $\beta (\lambda ,\kappa
) $ have been evaluated perturbatively employing (\ref{ab}) in section \ref%
{5}. In terms of these functions and combinations thereof defined in (\ref%
{ga1}) and (\ref{ga2}), the entries of the Hamiltonian (\ref{above}) are
given by, 
\begin{eqnarray}
h_{11}(\lambda ,\kappa ) &=&-(2{\gamma (\lambda ,\kappa )}^{4})^{-1}\left[
\alpha (\lambda ,\kappa )\beta (\lambda ,\kappa )\left( 6\alpha (\lambda
,\kappa )\beta (\lambda ,\kappa )-{\gamma (\lambda ,\kappa )}^{2}\right)
\right.  \notag \\
&&\left. +2\kappa \sinh (\gamma (\lambda ,\kappa ))\alpha (\lambda ,\kappa
)\gamma (\lambda ,\kappa )\delta (\lambda ,\kappa )\right.  \notag \\
&&\left. +2\cosh (\gamma (\lambda ,\kappa ))\delta (\lambda ,\kappa )\left(
2\lambda \alpha (\lambda ,\kappa )\beta (\lambda ,\kappa )+\delta (\lambda
,\kappa )\right) \right.  \notag \\
&&\left. +\lambda \alpha (\lambda ,\kappa )\left( {\alpha (\lambda ,\kappa )}%
^{2}-3\alpha (\lambda ,\kappa )\beta (\lambda ,\kappa )+4{\beta (\lambda
,\kappa )}^{2}\right) \epsilon (\lambda ,\kappa )\right.  \notag \\
&&\left. +\cosh (2\gamma (\lambda ,\kappa ))\alpha (\lambda ,\kappa )\left(
\beta (\lambda ,\kappa )-\lambda \alpha (\lambda ,\kappa )\right) {\epsilon
(\lambda ,\kappa )}^{2}\right.  \notag \\
&&\left. +\kappa \sinh (2\gamma (\lambda ,\kappa ))\beta (\lambda ,\kappa
)\gamma (\lambda ,\kappa ){\epsilon (\lambda ,\kappa )}^{2}\right] , \\
h_{22}(\lambda ,\kappa ) &=&\frac{\sinh (\gamma (\lambda ,\kappa ))}{\gamma
(\lambda ,\kappa )^{2}}\left[ \sinh (\gamma (\lambda ,\kappa ))\alpha
(\lambda ,\kappa )\left( \beta (\lambda ,\kappa )-\lambda \alpha (\lambda
,\kappa )\right) \right.  \notag \\
&&\left. \qquad \qquad \qquad +{\kappa }\cosh (\gamma (\lambda ,\kappa
))\beta (\lambda ,\kappa )\gamma (\lambda ,\kappa )\right] , \\
h_{44}(\lambda ,\kappa ) &=&({2{\gamma (\lambda ,\kappa )}^{4}})^{-1}\left[
\alpha (\lambda ,\kappa )\left( (6-\lambda )\alpha (\lambda ,\kappa ){\beta
(\lambda ,\kappa )}^{2}+\left( 1+4\lambda \right) {\beta (\lambda ,\kappa )}%
^{3}\right. \right.  \notag \\
&&\left. \left. +{\alpha (\lambda ,\kappa )}^{2}\beta (\lambda ,\kappa
)\left( 1-2\lambda \right) -\lambda {\alpha (\lambda ,\kappa )}^{3}\right)
+2\kappa \sinh (\gamma (\lambda ,\kappa ))\alpha (\lambda ,\kappa )\gamma
(\lambda ,\kappa )\delta (\lambda ,\kappa )\right.  \notag \\
&&\left. +2\cosh (\gamma (\lambda ,\kappa ))\delta (\lambda ,\kappa )\left(
2\lambda \alpha (\lambda ,\kappa )\beta (\lambda ,\kappa )+\delta (\lambda
,\kappa )\right) \right.  \notag \\
&&\left. +\cosh (2\gamma (\lambda ,\kappa ))\alpha (\lambda ,\kappa )\left(
\lambda \alpha (\lambda ,\kappa )-\beta (\lambda ,\kappa )\right) {\rho
(\lambda ,\kappa )}^{2}\right.  \notag \\
&&\left. -\kappa \sinh (2\gamma (\lambda ,\kappa ))\beta (\lambda ,\kappa
)\gamma (\lambda ,\kappa ){\rho (\lambda ,\kappa )}^{2}\right] , \\
h_{14}(\lambda ,\kappa ) &=&(2\gamma (\lambda ,\kappa )^{4})^{-1}\left[
-4\cosh (\gamma (\lambda ,\kappa ))\alpha (\lambda ,\kappa )\beta (\lambda
,\kappa )\left( 2\lambda \alpha (\lambda ,\kappa )\beta (\lambda ,\kappa
)+\delta (\lambda ,\kappa )\right) \right.  \notag \\
&&\left. +2\kappa \sinh (\gamma (\lambda ,\kappa ))\beta (\lambda ,\kappa
)\gamma (\lambda ,\kappa )\left( -2{\alpha (\lambda ,\kappa )}^{2}+\cosh
(\gamma (\lambda ,\kappa ))\delta (\lambda ,\kappa )\right) \right.  \notag
\\
&&\left. -\rho (\lambda ,\kappa )\left( \lambda {\alpha (\lambda ,\kappa )}%
^{2}-3\alpha (\lambda ,\kappa )\beta (\lambda ,\kappa )-2\lambda {\beta
(\lambda ,\kappa )}^{2}\right) \epsilon (\lambda ,\kappa )\right.  \notag \\
&&\left. -\cosh (2\gamma (\lambda ,\kappa ))\alpha (\lambda ,\kappa )\left(
\lambda \alpha (\lambda ,\kappa )-\beta (\lambda ,\kappa )\right) \epsilon
(\lambda ,\kappa )\rho (\lambda ,\kappa )\right]
\end{eqnarray}


\begin{thebibliography}{99}
\bibitem{YL} C.-N.~Yang and T.~D.~Lee, \newblock Statistical theory of
equations of state and phase transitions. I: Theory of condensation, %
\newblock Phys. Rev. \textbf{87}, 404--409 (1952).

\bibitem{LY} T.~D.~Lee and C.-N.~Yang, \newblock Statistical theory of
equations of state and phase transitions. II: Lattice gas and Ising model, %
\newblock Phys. Rev. \textbf{87}, 410--419 (1952).

\bibitem{Fisher} M.~E.~Fisher, \newblock Yang-Lee Edge Singularity and $%
\phi^3$ Field Theory, \newblock Phys. Rev. Lett. \textbf{40}, 1610--1613
(1978).

\bibitem{Cardy:1985yy} J.~L.~Cardy, \newblock Conformal invariance and the
Yang-Lee edge singularity in two-dimension, \newblock Phys. Rev. Lett. 
\textbf{54}, 1354--1356 (1985).

\bibitem{BPZ} A.~A.~Belavin, A.~M.~Polyakov, and A.~B.~Zamolodchikov, %
\newblock Infinite conformal symmetry in two-dimensional quantum field
theory, \newblock Nucl. Phys. \textbf{B241}, 333--380 (1984).

\bibitem{gehlen1} G.~von~Gehlen, 
\newblock {Critical and off critical conformal analysis of the Ising quantum
  chain in an imaginary field}, \newblock J. Phys. \textbf{A24}, 5371--5400
(1991).

\bibitem{gehlen2} G.~von~Gehlen, 
\newblock {NonHermitian tricriticality in the Blume-Capel model with imaginary
  field}, \newblock Int. J. Mod. Phys. \textbf{B8}, 3507--3529 (1994).

\bibitem{BB} C.~M.~Bender and S.~Boettcher, \newblock Real Spectra in
Non-Hermitian Hamiltonians Having PT Symmetry, \newblock Phys. Rev. Lett. 
\textbf{80}, 5243--5246 (1998).

\bibitem{Rev1} C.~Figueira~de Morisson~Faria, A.~Fring, and R.~Schrader, %
\newblock Analytical treatment of stabilization, \newblock Laser Physics 
\textbf{9}, 379--387 (1999).

\bibitem{Rev2} C.~M.~Bender, \newblock Making sense of non-Hermitian
Hamiltonians, \newblock Rept. Prog. Phys. \textbf{70}, 947--1018 (2007).

\bibitem{Rev3} A.~Mostafazadeh, \newblock Pseudo-Hermitian Quantum
Mechanics, \newblock arXiv:0810.5643.

\bibitem{EW} E.~Wigner, \newblock Normal form of antiunitary operators, %
\newblock J. Math. Phys. \textbf{1}, 409--413 (1960).

\bibitem{Bender:1998ke} C.~M.~Bender and S.~Boettcher, \newblock Real
Spectra in Non-Hermitian Hamiltonians Having PT Symmetry, \newblock Phys.
Rev. Lett. \textbf{80}, 5243--5246 (1998).

\bibitem{Bender:2002vv} C.~M.~Bender, D.~C.~Brody, and H.~F.~Jones, %
\newblock Complex Extension of Quantum Mechanics, \newblock Phys. Rev. Lett. 
\textbf{89}, 270401(4) (2002).

\bibitem{SW} S.~Weigert, \newblock $\mathcal{PT}$-symmetry and its
spontaneous breakdown explained by anti-linearity, \newblock J. Phys. 
\textbf{B5}, S416--S419 (2003).

\bibitem{Bender:2004sa} C.~M.~Bender, D.~C.~Brody, and H.~F.~Jones, %
\newblock Extension of PT-symmetric quantum mechanics to quantum field
theory with cubic interaction, \newblock Phys. Rev. \textbf{D70}, 025001(19)
(2004).

\bibitem{Korff:2007qg} C.~Korff and R.~A.~Weston, 
\newblock {PT Symmetry on the Lattice: The Quantum Group Invariant XXZ
  Spin-Chain}, \newblock J. Phys. \textbf{A40}, 8845--8872 (2007).

\bibitem{CKPT} C.~Korff, \newblock PT Symmetry of the non-Hermitian XX
Spin-Chain: Non-local Bulk Interaction from Complex Boundary Fields, %
\newblock J. Phys. \textbf{A41}, 295206 (2008).

\bibitem{Chico} F.~C.~Alcaraz, M.~N.~Barber, M.~T.~Batchelor, R.~J.~Baxter,
and G.~R.~W.~Quispel, \newblock Surface Exponents of the Quantum XXZ,
Ashkin-Teller and Potts Models, \newblock J. Phys. \textbf{A20}, 6397--6409
(1987).

\bibitem{DeGosh} T.~Deguchi and P.~Ghosh, \newblock Exactly Solvable
Quasi-hermitian Transverse Ising Model, \newblock arXiv:0904.2852.

\bibitem{Urubu} F.~G.~Scholtz, H.~B.~Geyer, and F.~Hahne, \newblock %
Quasi-Hermitian Operators in Quantum Mechanics and the Variational
Principle, \newblock Ann. Phys. \textbf{213}, 74--101 (1992).

\bibitem{Dieu} J.~Dieudonn{\'{e}}, \newblock Quasi-hermitian operators, %
\newblock Proceedings of the International Symposium on Linear Spaces,
Jerusalem 1960, Pergamon, Oxford , 115--122 (1961).

\bibitem{pseudo1} M.~Froissart, \newblock Covariant formalism of a field
with indefinite metric, \newblock Il Nuovo Cimento \textbf{14}, 197--204
(1959).

\bibitem{pseudo2} E.~C.~G.~Sudarshan, \newblock Quantum Mechanical Systems
with Indefinite Metric. I, \newblock Phys. Rev. \textbf{123}, 2183--2193
(1961).

\bibitem{Mostafazadeh:2001nr} A.~Mostafazadeh, \newblock Pseudo-Hermiticity
versus PT-Symmetry II: A complete characterization of non-Hermitian
Hamiltonians with a real spectrum, \newblock J. Math. Phys. \textbf{43},
2814--2816 (2002).

\bibitem{Dyson} F.~J.~Dyson, \newblock Thermodynamic Behavior of an Ideal
Ferromagnet, \newblock Phys. Rev. \textbf{102}, 1230--1244 (1956).

\bibitem{Holstein} T.~Holstein and H.~Primakoff, \newblock Field dependence
of the intrinsic domain magnetization of a ferromagnet, \newblock Phys. Rev. 
\textbf{58}, 1098--1113 (1940).

\bibitem{CA} C.~Figueira~de~Morisson~Faria and A.~Fring, \newblock Time
evolution of non-Hermitian Hamiltonian systems, \newblock J. Phys. \textbf{%
A39}, 9269--9289 (2006).

\bibitem{Mosta} A.~Mostafazadeh, 
\newblock {PT-symmetric cubic anharmonic
oscilator as a physical model}, \newblock J. Phys. \textbf{A38}, 6557--6570
(2005).

\bibitem{ACIso} C.~Figueira~de~Morisson~Faria and A.~Fring, \newblock %
Isospectral Hamiltonians from Moyal products, \newblock Czech. J. Phys. 
\textbf{56}, 899--908 (2006).

\bibitem{Can} E.~Caliceti, F.~Cannata, and S.~Graffi, \newblock Perturbation
theory of $\mathcal{PT}$-symmetric Hamiltonians, \newblock J. Phys. \textbf{%
A39}, 10019--10027 (2006).

\bibitem{AC1} A.~Fring, V.~Kostrykin, and R.~Schrader, \newblock On the
absence of bound-state stabilization through short ultra-intense fields, %
\newblock J. Phys. \textbf{B29}, 5651--567 (1996).

\bibitem{MGH} D.~P.~Musumbu, H.~B.~Geyer, and W.~D.~Heiss, \newblock Choice
of a metric for the non-Hermitian oscillator, \newblock J. Phys. \textbf{A40}%
, F75--F80 (2007).

\bibitem{Mostsyme} A.~Mostafazadeh, \newblock Metric operators for
quasi-Hermitian Hamiltonians and symmetries of equivalent Hermitian
Hamiltonians, \newblock J. Phys. \textbf{A41}, 055304 (2008).

\bibitem{PEGAAF} P.~E.~G.~Assis and A.~Fring, \newblock Metrics and
isospectral partners for the most generic cubic $\mathcal{PT}$-symmetric
non-Hermitian Hamiltonian, \newblock J. Phys. \textbf{A41}, 244001 (2008).

\bibitem{PEGAAF2} P.~E.~G.~Assis and A.~Fring, \newblock Non-Hermitian
Hamiltonians of Lie algebraic type, \newblock J. Phys. \textbf{A42}, 015203
(2009).

\bibitem{bendernew} C.~M.~Bender and S.~P.~Klevansky, \newblock Nonunique C
operator in PT Quantum Mechanics, \newblock arXiv0905.4673 .

\bibitem{kleefeld} F.~Kleefeld, \newblock The construction of a general
inner product in non- Hermitian quantum theory and some explanation for the
nonuniqueness of the C operator in PT quantum mechanics, \newblock %
arXiv0906.1011 .

\bibitem{baxter} R.~Baxter, \newblock Exactly solved models in statistical
mechanics, \newblock Academic Press , London--New York (1982).

\bibitem{izergin} V.~Korepin, N.~Bogoliubov, and A.~Izergin, \newblock %
Quantum inverse scattering method and correlation functions, \newblock %
Cambridge University Press (1993).

\bibitem{vNW} J.~von~Neuman and E.~Wigner, \newblock {\"U}ber merkw{\"u}%
rdige diskrete Eigenwerte. {\"U}ber das Verhalten von Eigenwerten bei
adiabatischen Prozessen, \newblock Zeit. der Physik \textbf{30}, 467--470
(1929).
\end{thebibliography}

\end{document}